\documentclass[final,3p,times,twocolumn]{elsarticle}

\usepackage{bm}
\usepackage{amssymb}
\usepackage{amsmath}
\usepackage{listings}
\usepackage{xspace}
\usepackage{color}  
\usepackage{pdflscape}  
\usepackage{rotating}  

\graphicspath{{figs/}}

\newcommand{\dA}{\mathrm{d}A}

\newcommand{\bu}{{\mathbf u}}
\newcommand{\br}{{\mathbf r}}
\newcommand{\bT}{{\mathbf T}}
\newcommand{\bS}{{\mathbf S}}
\newcommand{\bc}{{\mathbf c}}

\newcommand{\bn}{{\mathbf n}}

\newcommand{\numbat}{\textsf{NumBAT}\xspace}
\journal{Computer Physics Communications}

\begin{document}
\lstset{language=Python, basicstyle=\ttfamily}

\begin{frontmatter}

\title{\numbat: The integrated, open source Numerical Brillouin Analysis Tool}

\author[label1,label2]{Bj\"{o}rn C. P. Sturmberg}
\author[label3,label2]{Kokou B. Dossou}
\author[label3,label2,label4]{Michael J. A. Smith}
\author[label4,label2]{Blair Morrison}
\author[label3,label2]{Christopher G. Poulton}
\author[label1,label2]{Michael J. Steel}
\ead{michael.steel@mq.edu.au}
\address[label1]{MQ Photonics Research Centre,
Department of Physics and Astronomy, Macquarie University, NSW 2109, Australia}
\address[label2]{Centre for Ultrahigh-bandwidth Devices for Optical Systems (CUDOS)}
\address[label3]{School of Mathematical Sciences, University of Technology Sydney, Sydney, 2007, Australia}
\address[label4]{Institute of Photonics and Optical Science (IPOS),   School of Physics, The University of Sydney, NSW 2006, Australia}

\begin{abstract}
We describe \numbat, an open-source software tool for modelling stimulated Brillouin scattering in waveguides of arbitrary cross-section.
It provides rapid calculation of optical and elastic dispersion relations, field profiles and gain with an easy-to-use Python front end. 
Additionally, we provide an open and extensible set of standard problems and reference materials to facilitate the bench-marking of \numbat against subsequent tools. Such a resource is needed to help settle discrepancies between existing formulations and implementations, and to facilitate comparison between results in the literature. The resulting standardised testing framework will allow the community to gain confidence in new algorithms and will provide a common tool for the comparison of experimental 
designs of opto-acoustic waveguides.
\end{abstract}

\end{frontmatter}

\section{Introduction}
\label{intro}

The rapid growth of interest in stimulated Brillouin scattering (SBS) in integrated waveguides 
~\cite{Eggleton2013,Rakich:PRX:2012,VanLaer2015} has generated a need for new numerical tools for the analysis of integrated devices and their opto-elastic coupling.\footnote{The stimulated Brillouin literature frequently refers to both ``acoustic'' and ``elastic'' waves interchangeably. Here we favour the word ``elastic'' which carries a stronger
connotation of waves in solids. } The complex geometries required for the mutual confinement of light and sound, the sensitive dependence on waveguide materials and configuration, as well as 
the possibility of strongly anisotropic elastic properties, calls for advanced and precise tools
that meet the requirements of experimental design and interpretation. 
Currently researchers rely on a mix of commercial and in-house software, including  bespoke 
routines that vary substantially between research groups. The variety of formulations and implementations,
together with the intricacy of the problem and the absence of standardised tests, 
makes it challenging to reproduce existing results and thereby to validate one's own calculations.
This task is made more difficult by the fact that in certain waveguide geometries and SBS configurations, some effects 
contribute very little or cancel out, but dominate when a different geometry or configuration is used. 
This means that errors either in the underlying equations or in the implementation can be well hidden,
and there is currently no universally-accepted simulation package that can be used to validate new results. 
Moreover, even assuming identical formulations and implementations of the opto-elastic waveguide problem,
there is substantial literature variation in the numerical values of material properties, both elastic and optical, 
again making tests against earlier results difficult. Indeed, for many contemporary 
materials such as soft glasses, the elastic properties have never been reliably measured at all.
The community would benefit greatly from an open library of reference materials that can be continually expanded.


We have developed  an open-source and freely available simulation package named \numbat, the \emph{Numerical Brillouin Analysis Tool}, that is specifically designed to carry out SBS calculations for longitudinally invariant waveguides. The tool calculates both forward and backward Brillouin gain spectra
for both inter- and intra-mode scattering, as well as providing the optical and elastic modes. Other key quantities needed for the design of SBS-active waveguides and for the interpretation of experiments are also provided, such as elastic quality factors. The implementation accounts for both electrostrictive/photo-elastic and radiation pressure/moving boundary 
effects, with the roto-optic contribution~\cite{Smith2017} to be added shortly. The code uses a fully-tensorial formulation for the computation of  the mechanical problem, and so anisotropic materials can be examined. A sub-class of optically anisotropic materials is also supported.  The code can also be used to compute the elastic wave loss if the material components of the viscosity tensor are known. The package automatically identifies the symmetry class of optical and elastic fields to help identify suitable mode combinations for efficient opto-elastic coupling~\cite{Wolff2014}
and to facilitate understanding of the complex families of elastic dispersion relations and crossings 
that are typically found in these calculations.

\numbat builds specifically on the mathematical formalism developed by Wolff {\em et al.}~\cite{Wolff2015},
which was designed to take into account the tensorial nature of the opto-mechanical interaction, 
as well as to avoid questions involving the interpretation of optical momenta in materials.  
In preparation for this article we have confirmed the compatibility of the formalism of Wolff {\em et al.} 
with others in the literature, including those of Rakich {\em et al.}~\cite{Rakich:PRX:2012},  Florez {\em et al.}~\cite{Florez:NC:2016}
 and Sipe and Steel~\cite{Sipe2016}, and have clarified implicit assumptions or corrected minor errors 
that exist in all these works (including our own~\cite{Wolff2015,Sipe2016}). 
In this paper we present the corrected formalism, discuss various pitfalls that can arise in both the analysis and
in the numerics, and focus on the open-source implementation of the resulting algorithm.

The code is freely available on GitHub~\cite{numbatcode}, depends solely on open-source languages and libraries, and comes with extensive documentation and tutorials~\cite{numbatdocs},  largely based on experimental and theoretical results drawn from the literature. \numbat's theoretical formulation, numerical implementation, material parameters and simulation results can all be transparently inspected and cross-checked, thereby allowing users to validate the results. Access to \numbat's internal results at different stages of the calculation  allows authors of new tools to benchmark their software, so that the community can gain confidence in new algorithms as rapidly as possible. The authors welcome the contribution of extensions from the community under a standard open source license. 
The GitHub repository also contains a materials library that allows users to 
submit their own material models for use by others. 

\numbat is built on two fast Fortran-based two-dimensional Finite Element Method (FEM) routines that solve for the optical and elastic modes of arbitrary longitudinally invariant waveguides.
The FEM presented in~\cite{Dossou:CMAME:2005} has been used for the computation of the optical modes.
This routine also underlies a recent tool presented by some of us for electromagnetic scattering in complex layered structures~\cite{Sturmberg:CPC:2016}. 
The numerical implementation of the FEM, for the elastic modes of a waveguide, is similar to the one described in~\cite{Hladky:Hennion:JSV:1996}.
All materials are assumed to be optically linear and non-magnetic, and while their elastic properties may be fully anisotropic, their permittivity tensor must (currently) be at most uniaxially anisotropic.
The numerically intensive evaluation of the opto-elastic interaction integrals is also implemented in optimised Fortran routines. 

Geometries and materials are specified in a user-friendly class-based Python front-end with classes representing the familiar tensors or elastic moduli and related parameters. The Python interface provides access to all values and geometries required by typical users, and enables the use of the extensive suite of Python libraries, such as \textsf{numpy}, \textsf{scipy} and \textsf{matplotlib}, for convenient data analysis and multiprocessing.

The remainder of this paper is organised as follows: in Sect.~\ref{sec:approach} we outline the theoretical treatment of SBS including the key expressions, in Sect.~\ref{sec:implement} we describe the formulation's implementation in \numbat, and in Sect.~\ref{sec:tests} we present simulation examples which may serve as standard reference SBS calculations. 
In~\ref{sec:FEM} we present the FEM formulation applied to the problem of solving for the elastic modes of a waveguide. 


\section{Approach Outline}
\label{sec:approach}

To study Brillouin scattering we require descriptions of optical and elastic propagation and
the interaction between both fields.  We consider the class of waveguide problems where both families of waves propagate along a common axis $z$ and are, at least weakly, confined within the plane perpendicular to this axis.
In this case the calculation of the Brillouin scattering interaction involves three main steps: 
\begin{enumerate} 
  \item solving for the optical modes (fields and wavevectors) of the waveguide at a given frequency,
  \item solving for the elastic modes (fields and frequencies) of the waveguide at a given wavevector (generally set relative to the wavevector of a chosen optical mode),
  \item evaluating the overlap integrals of the optical and elastic modal fields that describe the interaction between the fields.
\end{enumerate}
The approach taken by \numbat is to allow the user to specify the two optical modes, and then compute the
relevant elastic mode from phase-matching requirements (see Section \ref{sec:modal_fields}). 
This approach maximises the flexibility of the computation: the modes may be co- or counter-propagating, 
of different polarisations or mode order, or may even differ significantly in frequency.



The geometry that we consider is that of a $z$-invariant waveguide capable of supporting both optical and acoustic waveguide modes. The electromagnetic properties of such a structure are described by the relative permittivity $\varepsilon_r (x,y)$, with the acoustic properties given by the components of the stiffness tensor $C_{ijkl} (x,y)$ and the density $\rho(x,y)$ (see Appendix B). The coupling between optical and acoustic waves depends on the photo-elastic effect, described by the symmetric fourth rank tensor $p_{ijkl} (x,y)$. All elastic properties are defined across the cross-section of the waveguide; a vacuum condition (for an ideal suspended waveguide) can be modelled by imposing stress-free boundary conditions (see Appendix B). 

For the guided electromagnetic modes, each constituent mode, labelled $n$, 
of the pump (p) and Stokes (s) waves is taken to be of the form
\begin{align}
  \mathbf{E}_n^{\rm (p,s)}(\mathbf{r},t) &= \mathbf{e}_n^{\rm (p,s)}(x,y)\, \textrm{exp}(i k_n^{\rm (p,s)} z - i \omega_n^{\rm (p,s)} t) + \mbox{c.c.}, \\
  \mathbf{H}_n^{\rm (p,s)}(\mathbf{r},t) &= \mathbf{h}_n^{\rm (p,s)}(x,y)\, \textrm{exp}(i k_n^{\rm (p,s)} z - i \omega_n^{\rm (p,s)} t) + \mbox{c.c.}, 
\end{align}
where $\mathbf{e}_n^{\rm (p,s)}(x,y)$ and $\mathbf{h}_n^{\rm (p,s)}(x,y)$ are the electric and magnetic parts of the modal field distribution, $k_n^{\rm (p,s)}$ is the propagation constant in the positive $z$-direction, $\omega_n^{\rm (p,s)}$ is the optical angular frequency, $t$ is time and c.c. represents the complex conjugate. Each constituent mode of the elastic wave, labelled $m$, is taken to be of the form 
\begin{equation}
  \mathbf{U}_m(\mathbf{r},t) = \mathbf{u}_m(x,y)\, \textrm{exp}(i q_m z - i \Omega_m t) + \mbox{c.c.},
\end{equation}
where $\mathbf{u}_m(x,y)$ is the modal field distribution, $q_m$ is the propagation constant in the positive $z$-direction and $\Omega_m$ is the elastic angular frequency. We emphasise that these modes are all fully vectorial, and include the longitudinal  components $e_z$, $h_z$ and $u_z$.

\subsection{Modal solutions}
The electromagnetic modes are solutions to the vector Helmholtz equation which follows from Maxwell's equations. In \numbat, we solve the Helmholtz equation in the form 
\begin{equation}
-\nabla \times \left( \nabla \times \mathbf{E} \right) + \omega^2 \varepsilon_0 \varepsilon_r \mathbf{E} = \mathbf{0}
\end{equation}
with the $\mathbf{h}$ field found from Ampere's/Faraday's law. 

The elastic modes are solutions to the elastic wave equation 
\begin{equation}
\nabla \cdot \mathbf{T} + \Omega^2 \rho \, \mathbf{u}= \mathbf{0}.
\end{equation}
Here $\mathbf{T}$ is the elastic stress tensor, which is related to the linear strain tensor $\mathbf{S}$ by Hooke's law $\mathbf{T}=\mathbf{c} : \mathbf{S}$.
Here $\mathbf{c}$ is the spatially-varying fourth-rank \emph{stiffness tensor} which describes the elastic material properties of the waveguide.
A fuller presentation of the relevant elastic theory and our approach to constructing a finite element formulation is provided in \ref{sec:FEM}.

\subsection{Modal properties}
The power carried in each optical mode $n$ (calculated in Watts in \numbat) is given by
\begin{equation}
\label{P_opt}
  P_n^{\rm (p,s)} = 2 {\rm Re} \int {\rm d}^2r \, \hat{z} \cdot (\mathbf{e}_n^{\rm (p,s)*} \times \mathbf{h}_n^{\rm (p,s)}),
\end{equation}
where the integration is over the whole transverse plane.
The corresponding linear energy density of the optical mode in units J$\cdot$m$^{-1}$ is
\begin{equation}
\mathcal{E}_n^{(\rm p, \rm s)} = 2  \varepsilon_0 \int {\rm d}^2r \, \varepsilon_r(x,y) \, ||\mathbf{e}_n||^2 ~,
\end{equation}
where $\varepsilon_r$ is the (position-dependent) relative dielectric constant and $\varepsilon_0$ is the vacuum permittivity. 

The linear density of elastic energy in units $\mbox{J}\cdot\mbox{m}^{-1}$ for the elastic modes is given by~\cite{Sipe2016}
\begin{equation}
\label{E_ac}
  \mathcal{E}_m^{\rm (a)} = 2 \Omega^2 \int_A {\rm d}^2r \, \rho ||\mathbf{u}||^2, 
\end{equation}
where the integration is strictly over the cross section of the waveguide, 
which is treated as being surrounded by vacuum.
The corresponding modal elastic power is 
\begin{equation}
\label{P_ac}
P_m^{\rm (a)} = {\rm Re}  \int_A {\rm d}^2r \, (-2i \Omega) \sum_{jkl} c_{zjkl} {u}_{mj}^{*} \partial_k {u}_{ml} ~,
\end{equation}
where $c_{ijkl}$ is the position-dependent stiffness tensor of the waveguide.
Note that following~\cite{Wolff2015}, we do not normalise the energy in each mode, instead carrying the energy terms through to the gain calculation in Eq.~\eqref{gain}.

While in practice one must typically consider the SBS gain contributions of many elastic modes, as well as potentially different optical modes, in the interest of clarity we proceed with the treatment of an individual set of modes (one pump, one Stokes, one elastic), omitting the mode index subscripts and leaving the generalisation to Sect.~\ref{sec:implement}.

The SBS gain of a particular combination of pump, Stokes and elastic modes is calculated as follows:
\begin{enumerate}

\newcommand{\QPE}{Q^\text{PE}}
\item \emph{Calculate the photoelastic (PE) interaction strength, which is the inverse process to electrostriction.} 
This is given in Eq.~(33) 
in~\cite{Wolff2015},
\begin{equation}
\label{PE}
  \QPE = - \varepsilon_0 \int_A {\rm d}^2\mathbf{r} \sum_{ijkl} \varepsilon^2_r e^{\rm (s)*}_i e^{\rm (p)}_j p_{ijkl} \partial_k u_l^{*},
\end{equation}
where $\QPE$ is the photoelastic coupling in units $\mbox{W}\cdot \mbox{m}^{-1}\cdot\mbox{s}$ and $p_{ijkl}$ is the dimensionless photoelastic tensor.
\footnote
{Note that here we have corrected a sign error in Eq.~(31) and~(33) of~\cite{Wolff2015}, so that the 
photoelastic tensor here matches the standard convention that the electric susceptibility is 
$\chi^\text{PE}_{ij}=-\varepsilon_r^2 \sum_{kl} p_{ijkl}\partial_k u_l$.}

\item \emph{Calculate the moving boundary (MB) interaction, which is the inverse phenomenon to radiation pressure.} 
This coupling, measured in units $\mbox{W}\cdot \mbox{m}^{-1}\cdot\mbox{s}$, is given by~\cite{Rakich:PRX:2012}
\begin{equation}
\label{MB}
\begin{aligned}
  Q^{\rm MB} = \int_{\cal C} &{\rm d \mathbf{r} \, (\mathbf{u}^{*} \cdot \hat n}) \\
  & \big[ (\varepsilon_a - \varepsilon_b)  \varepsilon_0 ({\rm \hat n \times \mathbf{e^{\rm (s)}}})^{*} \cdot ({\rm \hat n \times \mathbf{e}^{\rm (p)}}) - \\ 
  &(\varepsilon_a^{-1} - \varepsilon_b^{-1})  \varepsilon_0^{-1} ({\rm \hat n \cdot \mathbf{d}^{\rm (s)}})^{*} \cdot ({\rm \hat n \cdot \mathbf{d}^{\rm (p)}}) \big],
\end{aligned}
\end{equation}
where the integral is taken in the positive sense with respect to the outward normal around the contour of the waveguide boundary
$\cal C$.
\footnote{There is a sign error in Eq.~(41) in~\cite{Wolff2015}, which is corrected here by the replacement $(\varepsilon_b^{-1} - \varepsilon_a^{-1}) \mapsto (\varepsilon_a^{-1} - \varepsilon_b^{-1})$. }

\item \emph{Finally, calculate the SBS gain.} The peak SBS gain, $\Gamma_0$, in units $\mbox{W}^{-1}\cdot\mbox{m}^{-1}$ of this particular combination of pump, Stokes and elastic modes is given by Eq.~(91) in~\cite{Wolff2015},
\begin{equation}
\label{gain}
  \Gamma_0 =  \frac{2\omega\Omega}{\alpha} \frac{ (|Q^{\rm Tot}|^2)}{P^{\rm (s)} P^{\rm (p)} \mathcal{E}^{\rm (a)}}, 
\end{equation}
where
\begin{equation}
\label{Q:Tot}
Q^{\rm Tot} = Q^{\rm PE} + Q^{\rm MB},
\end{equation}
$\alpha$ is the temporal elastic attenuation coefficient in units $\mbox{s}^{-1}$, and $\omega$ is chosen to be $\omega^{\rm (p)}$, which typically varies from $\omega^{\rm (s)}$ by less than $0.01\%$.
We note that, in contrast to~\cite{Wolff2015}, we normalise with the elastic elastic energy (Eq.~\eqref{E_ac} and Eq.~(16) in~\cite{Wolff2015}) rather than elastic elastic power (Eq.~18 in~\cite{Wolff2015}). This choice was made to better handle forward SBS, where the power flow associated with transverse elastic oscillations is near zero while the energy density remains finite. The elastic power integral (Eq.~\eqref{P_ac}) is also implemented in \numbat in case it is of interest to users. We also note that because the modal powers are defined in accordance with the direction of energy carried (negative power for a backward-propagating wave) the SBS gain arising from Eq.~\eqref{gain} will be positive for co-propagating interactions, and negative for counter-propagating or backward SBS. 
The gain driven by solely the photoelastic or radiation pressure effects is given by Eq.~\eqref{gain} with $Q^{\rm Tot} = Q^{\rm PE}$ or $Q^{\rm Tot} = Q^{\rm MB}$ respectively.

The elastic attenuation due to dissipation can be derived by considering the propagating of an elastic mode in a waveguide possessing a non-zero viscosity tensor (see p.~143 in~\cite{Auld:book:1973}). In terms of the visco-elastic tensor $\eta_{ijkl}$ and the deformation, the loss per unit time (\emph{i.e.}, in units ${\rm s}^{-1}$) of the elastic amplitude is
\begin{equation}
\label{alpha}
  \alpha = \frac{\Omega^2}{\mathcal{E}^{(a)}} \int_A {\rm d}^2r \sum_{ijkl} \partial_i u_j^{*} \eta_{ijkl} \partial_k u_l~.
\end{equation}
We note that a similar expression given in Eq.~(45) in~\cite{Wolff2015} neglects contributions
from the boundaries when performing an integration by parts of Eq.~\eqref{alpha}.

Further attenuation occurs due to scattering and radiation losses. We do not here present theoretical expressions for these losses, but their contributions can be included by calculating $\alpha$ based on an assumed quality factor $Q_{\rm loss}$, which may be estimated from experimental measurements. In such a case 
\begin{equation}
\label{alphaQ}
\alpha = \frac{v_g q}{2 Q_{\rm loss}},
\end{equation}
where $v_g$ is the group velocity of the elastic mode.
If, for example, the frequency and spectral linewidth of a mode's SBS resonance is known, then $Q_{\rm loss} = \Omega_0/(2\pi\gamma)$, where $\Omega_0$ is the central angular frequency, and $\gamma$ is the spectral linewidth with respect to angular frequency in units $\mbox{s}^{-1}$. 

It is often of interest to plot the SBS gain profile in frequency, including the contributions from multiple modes. To facilitate this, \numbat assumes each resonance to be Lorentzian with the linewidth $\gamma = \alpha/2\pi$,
\begin{equation}
  \label{gain2}
  \Gamma(\delta \Omega) =  \Gamma_0 \frac{(\gamma/2)^2}{(\gamma/2)^2 + (\delta \Omega)^2}, 
\end{equation}
where $\delta \Omega$ is the range of angular frequency detuning.
\end{enumerate}

Having established the fundamentals of the theoretical formulation, we now move on to the how this is implemented in \numbat.

\section{\numbat Implementation}
\label{sec:implement}
\numbat is implemented using a combination of Fortran and Python. Fortran is chosen for the numerically intensive subroutines because of its numerical efficiency and its easy access to extensively optimised open source linear algebra packages 
including BLAS, LAPACK 
and UMFPACK~\cite{UMFPACK:ACM:2004}.
The user-friendly scripting language Python is used for the remainder of the program,  providing full access to all variables, and thereby facilitating both high level parameter sweeps and low level manipulation of basic field quantities. The communication between subroutines written in each language is handled by the f2py package~\cite{Peterson:IJCSE:2009}, which creates a Python wrapper around the Fortran source code, with pointers to memory blocks being passed directly to Python creating minimal computational overhead. In this way \numbat balances the need for efficient numerical computation and ease of use.
Following the Unix philosophy to: ``Write programs that do one thing and do it well'', \numbat comes without a graphical user interface, opting instead for control through Python script files.

\numbat uses frequency domain solvers, where each frequency is independent, and problems that span across wavelength ranges are `embarrassingly parallel'. This means that the calculation at each wavelength may run in parallel on separate CPUs without any communication between them. This is done in Python using the {\ttfamily multiprocessing} package. To minimise the computational overhead and thereby maximise the speed enhancement of parallelisation we do not introduce any parallel elements to the calculation for each wavelength, though users may use parallel versions of the linear algebra packages. Calculations across structural parameters are also embarrassingly parallel, and the script file approach is particularly well suited to setting up a scan of multiple parameters to be carried out in parallel.

\begin{figure*}
\centerline{
   \includegraphics[width=0.8\textwidth]{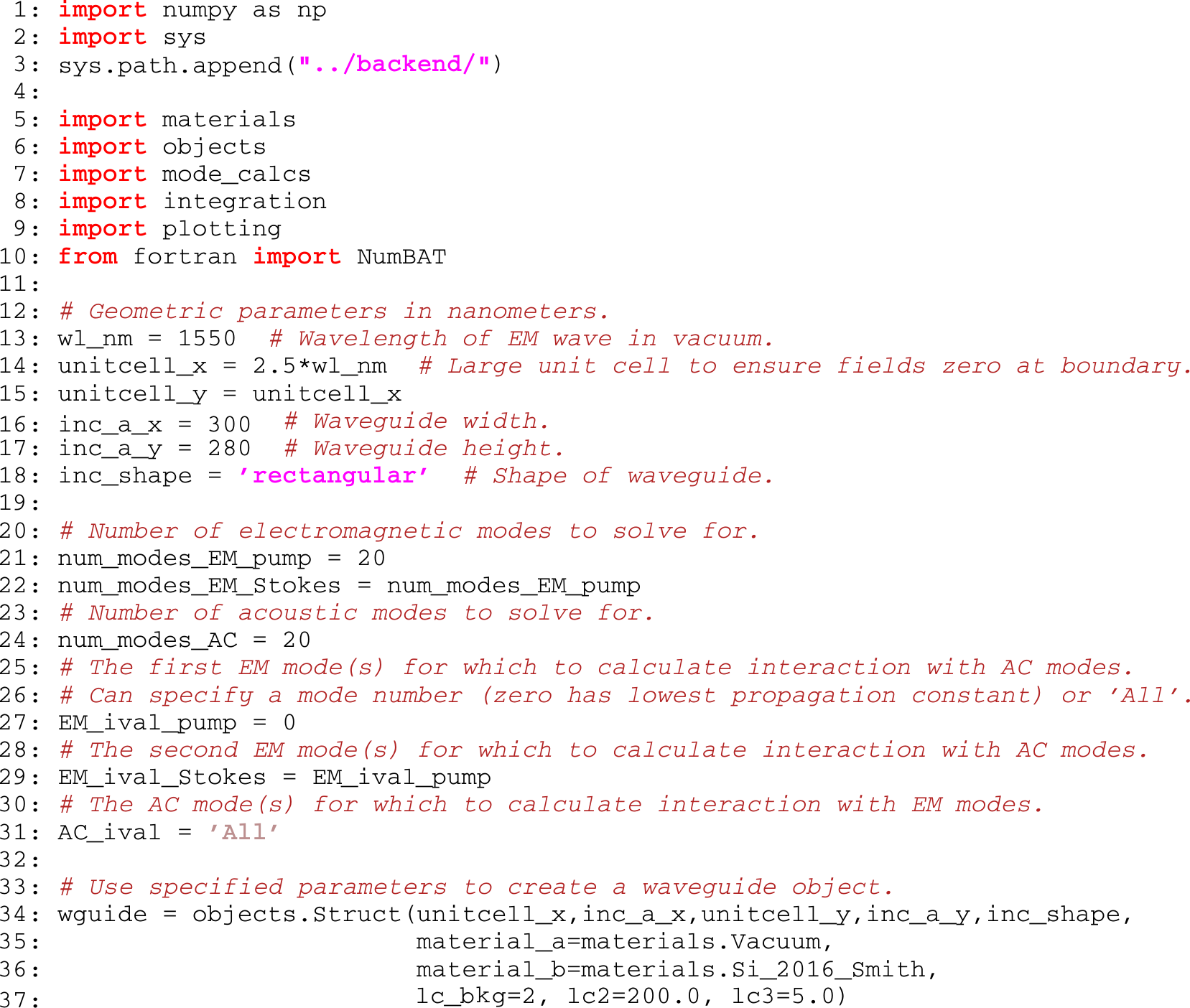}
}
\caption{Example simulation part 1/3, showing the specification of parameters and the initialisation of the waveguide object.}
\label{fig:simo}
\end{figure*}

To illustrate the approach of \numbat we present an example simulation, taken from the included tutorial suite, which is shown in Figs.~\ref{fig:simo}-\ref{fig:simo3}.


\subsection{Setup}
Figures~\ref{fig:simo}--\ref{fig:simo3} show a basic \numbat control script.
As seen in Fig.~\ref{fig:simo},  a \numbat simulation begins with the specification of the optical wavelength and the waveguide's geometry and constituent materials. In the example of Fig.~\ref{fig:simo}, the optical wavelength is set in line 12, the key parameters of the waveguide are defined in lines 13-18 (the prefix {\ttfamily inc} stands for ``inclusion''), and the waveguide object is initialised in lines 34-37 where the FEM mesh is created using the open-source program Gmsh~\cite{Gmsh:IJNME:2009}.
In lines 21-31 we specify the desired number of modes, and which combinations of these modes will be included in the gain calculations. 

In this example simulation we consider backward SBS, where the pump and Stokes waves both occupy the fundamental mode of the waveguide. In this case we seek results for the first ten eigenmodes, and specify the minimum number of modes in the ARPACK expansion required for stability at twice this number with {\ttfamily num\_modes\_EM\_pump = 20}. We then specify that only the modal combinations including the fundamental mode {\ttfamily EM\_ival\_pump = 0}, {\ttfamily EM\_ival\_stokes = 0} need to be considered in the calculation of the gain, which saves considerable computation time.
For the same reason, the number of elastic modes must also be at least 20. Values greater than this are recommended when a wide gain spectrum is desired.

\subsection{Modal Fields}
\label{sec:modal_fields}
To initialise the search for the modes of the waveguide we must supply the FEM modesolver with an estimate of the wavevector of the target mode, which is most conveniently achieved by estimating the mode's effective index ({\ttfamily n\_eff}). In the example simulation,  we are interested in the fundamental mode and estimate its effective index as being slightly less than the refractive index of the waveguide material (Fig.~\ref{fig:simo2}, line 40). The mode solver is relatively insensitive to this initial estimate. Calculating the optical modes of the pump wave is then a simple one line call (Fig.~\ref{fig:simo2}, line 42), which returns a Python object that contains the wavevectors (eigenvalues), fields (eigenvectors), and optical power (as per Eq.~\eqref{P_opt}) of each mode.

In the example we consider a simplified backward SBS computation, in which we observe that 
the difference in frequency between pump and Stokes is sufficiently small, and the 
optical dispersion sufficiently low, that the 
Stokes modes are almost identical to backward-propagating pump modes.
In this calculation the {\ttfamily mode\_calcs.bkwd\_Stokes\_modes(sim\_EM\_pump)} routine transforms the pump modes, reversing the direction of a mode by conjugating its fields and reversing the wavevector. 

\begin{figure*}
\begin{center}
   \includegraphics[width=0.8\textwidth]{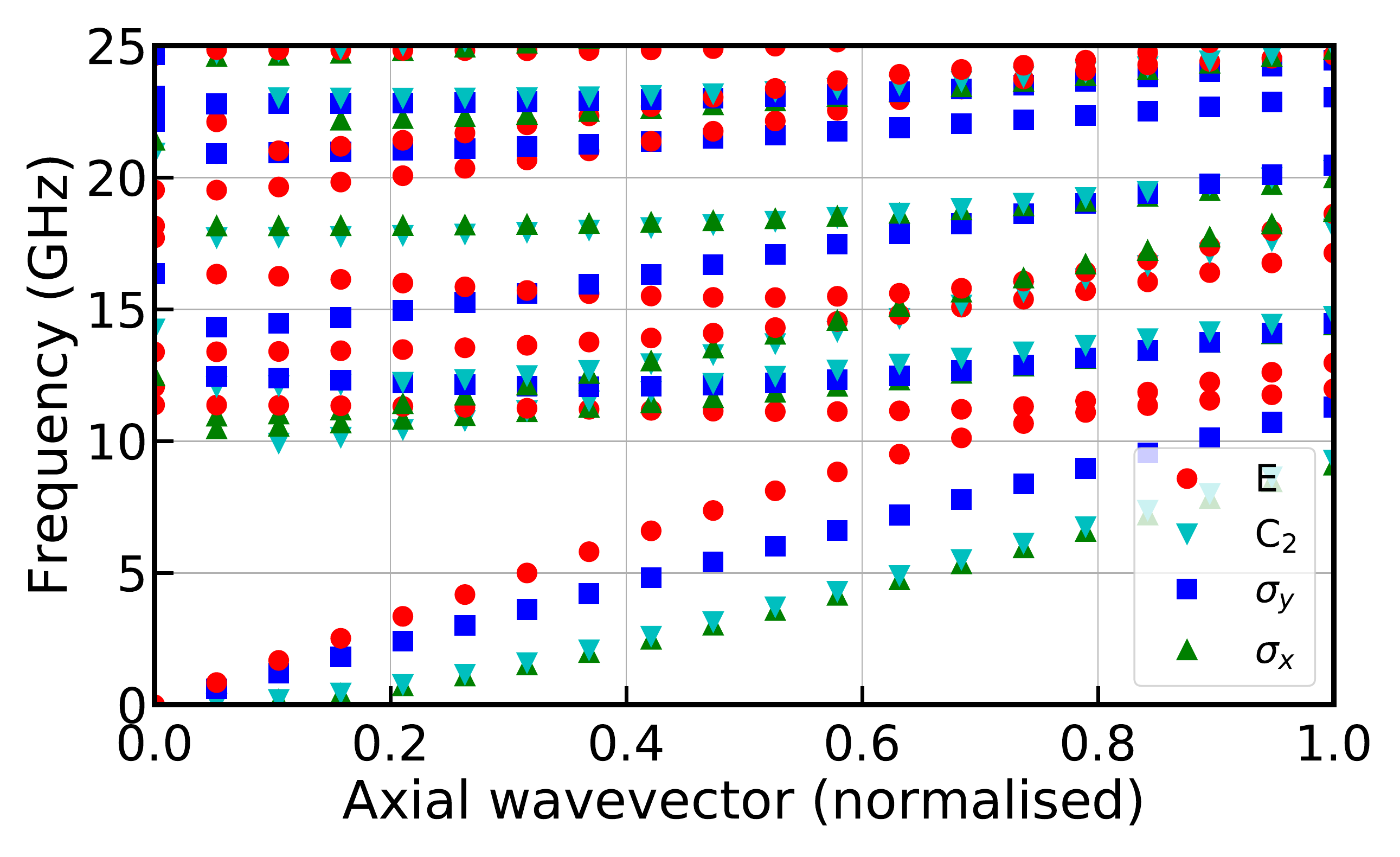}
\end{center}
\caption{Elastic dispersion relation for a geometry similar to that in Ref.~\protect\cite{Rakich:PRX:2012}, included
as Tutorial~3 in the \numbat documentation. The structure is a rectangular silicon waveguide of 
dimensions 314.7~nm by 283.2~nm.
The different colours allow identification of elastic modes of different symmetry.  }
\label{fig:elasticdr}
\end{figure*}

\begin{figure*}
\begin{center}
   \includegraphics[width=0.8\textwidth]{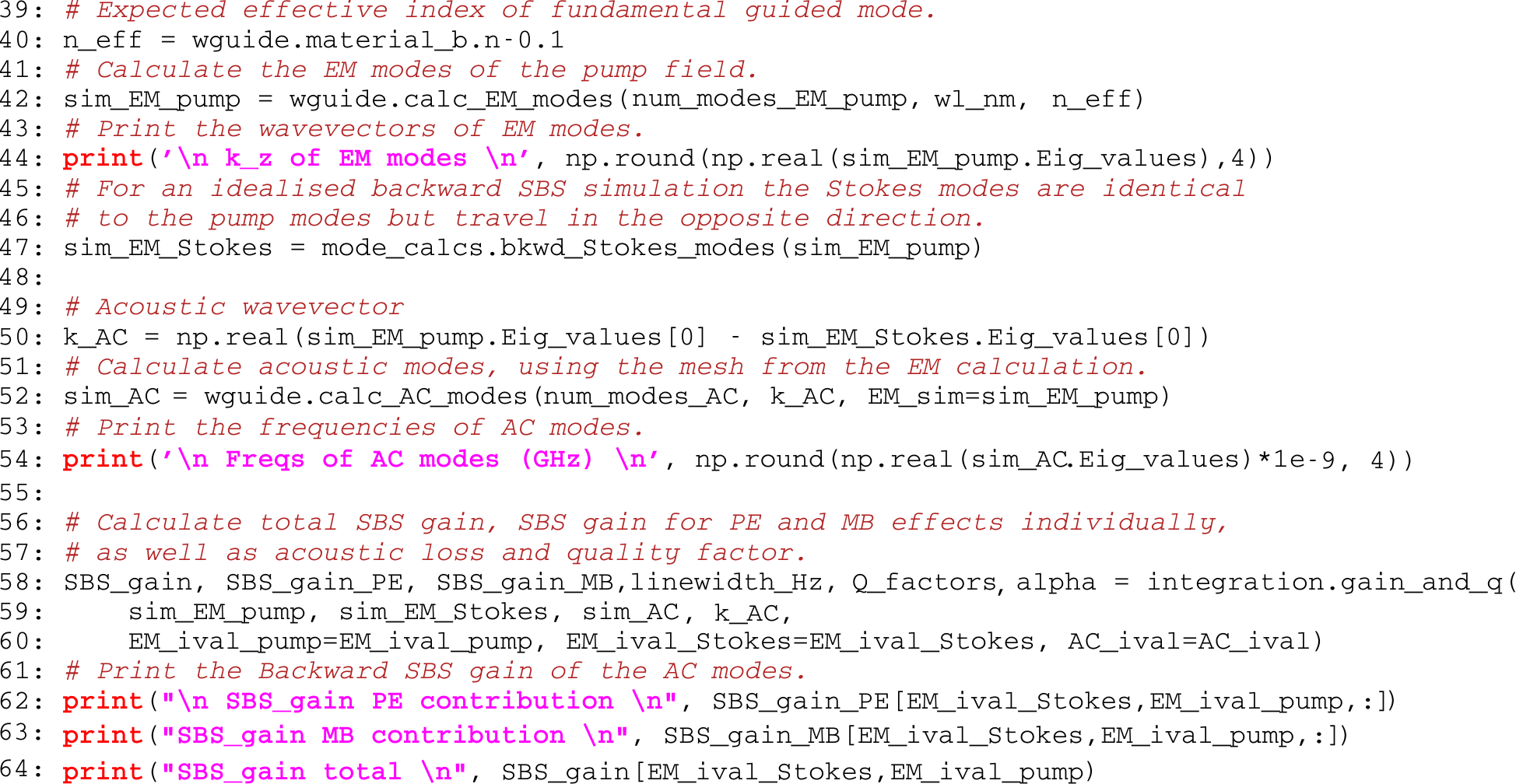}
\end{center}
\caption{Example simulation part 2/3, showing the calculation of the optical and elastic modes as well as the SBS gain.}
\label{fig:simo2}
\end{figure*}

In all cases, the wavevector of the elastic modes is equal to the difference in wavevectors of the pump and Stokes waves (Fig.~\ref{fig:simo2} line 50) 
\begin{equation}
\label{phase:matching:condition}
q = k^{\rm (p)} - k^{\rm (s)}.
\end{equation}
In the call to {\ttfamily wguide.calc\_AC\_modes} in line 52,
the elastic FEM modesolver takes this wavevector $q$ as an input and returns the frequencies $\Omega_m$ of the elastic modes. (Note that this is a reverse procedure to that  of the optical modesolver which solves for the wavevectors $k$ for given frequency $\omega$). The other key input into the elastic FEM is a modified version of the FEM mesh, where any regions representing vacuum have been removed. This is generally best done by passing through the Python object that contains the results of the optical simulation as the argument {\ttfamily sim\_EM=sim\_EM\_pump}, in which case \numbat handles the conversion. 
If users wish to compute just the elastic modal properties of the waveguide, they can create a FEM mesh that covers the waveguide cross-section, and leave out the keyword argument {\ttfamily sim\_EM}.
The FEM modesolver also requires an initial estimate of where to start its search, which in the elastic case is by default estimated in the {\ttfamily calc\_AC\_modes} routine, based on the group velocity of a longitudinal mode in bulk. The estimate of the lowest frequency of interest can be set using the keyword argument {\ttfamily shift\_Hz}.
The calculation of the elastic modes is a simple one line call (Fig.~\ref{fig:simo2} line 52), which returns a Python object that contains the frequencies (eigenvalues), fields (eigenvectors), and energy density (as per Eq.~\eqref{E_ac}) of each mode.
Figure~\ref{fig:elasticdr} shows a calculated dispersion relation for the elastic modes for a geometry similar to a rectangular waveguide studied in Ref.~\cite{Rakich:PRX:2012}.
The figure illustrates that where possible \numbat classifies the elastic modes by their symmetry class, which allows quick identification of modes that will yield a nonzero coupling between the desired optical modes~\cite{Wolff2014}.

We note that solving the eigensystems to find the optical and elastic modes is by far the most computationally expensive step in \numbat and that at the completion of these steps the SBS problem is now reduced to numerical quadrature.

\subsection{SBS Interactions}
The SBS gain may now be found by evaluating the SBS (photoelastic and moving boundary) interaction as described in the integrals of Eqs.~\eqref{PE}--\eqref{gain}. This is done in \numbat using the {\ttfamily integration.gain\_and\_qs} function as shown in lines 58-61 of Fig.~\ref{fig:simo2}.
This function returns the total peak SBS gain ($Q^{\rm Tot} = Q^{\rm PE} + Q^{\rm MB}$, {\ttfamily SBS\_gain}), as well as the peak SBS gain due solely to the photoelastic ($Q^{\rm Tot} = Q^{\rm PE}$, {\ttfamily SBS\_gain\_PE}) and moving boundary ($Q^{\rm Tot} = Q^{\rm MB}$, {\ttfamily SBS\_gain\_MB}) effects.
The function
{\ttfamily integration.gain\_and\_qs} also evaluates the elastic loss as per Eq.~\eqref{alpha}, unless predetermined quality factors are input using the {\ttfamily fixed\_Q} keyword argument, in which case $\alpha$ is set as per Eq.~\eqref{alphaQ}.

The evaluation of these integrals for all combinations of optical and elastic
modes is relatively computationally intensive and they are therefore
implemented in \numbat using Fortran subroutines.  The surface integrals of the
photoelastic interaction and the elastic loss (Eqs.~\eqref{PE} and
\eqref{alpha}, respectively) have been implemented twice: once using a
semi-analytical approach that is valid on rectilinear triangles (the integral over each triangle is computed analytically,) and may therefore be applied on purely rectangular structures, 
while the second subroutines use
numerical quadrature over each triangle and can therefore be applied on any mesh, including ones
that contain curvi-linear triangles. The (one-dimensional) contour integral of
Eq.~\eqref{MB} (moving boundary effect) is also implemented in a Fortran
subroutine. This subroutine identifies elements that form the boundary between
different materials, orientates the edges of these mesh elements consistently
to form a contour, and then evaluates the integral using numerical quadrature.

\subsection{Plotting Fields and Gains} 
\begin{figure*}
\begin{center}
   \includegraphics[width=0.8\textwidth]{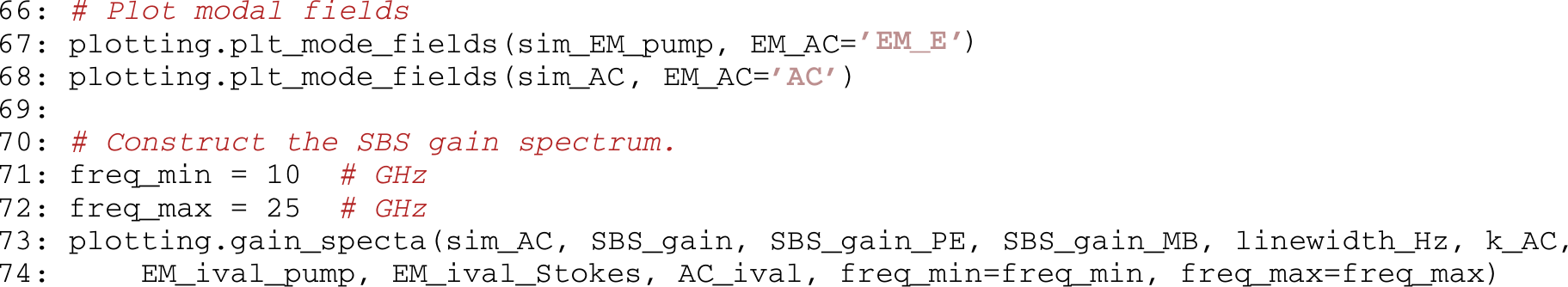}
\end{center}
\caption{Example simulation part 3/3, showing the plotting of modal fields and SBS gain spectrum.}
\label{fig:simo3}
\end{figure*}

Lastly, we plot the fields of the
optical and elastic modes as well as the SBS gain spectrum. This is done in
lines 67,~68 and 71-74 of Fig.~\ref{fig:simo3}. The gain spectrum is the
superposition of the Lorentzian resonances of each combination of the optical
and elastic modes, calculated using Eq.~\ref{gain2}.
The {\ttfamily plotting.gain\_spectra} function generates a plot of the gain spectrum due to the photoelastic effect, the gain spectrum due to the moving
boundary effect, and the total gain spectrum (as shown for example in
Fig.~\ref{fig:lit_04-no_pillar-gain_spectra-MB_PE_comps}). The same function can be instructed to plot the Lorentzian gain curves of each combination of individual optical and acoustic modes.

\section{Standard Benchmarks}
\label{sec:tests}

One of our major motivations in developing \numbat was to create a set of open source reference studies that the research community can use to benchmark other software. For such a database to be trustworthy it is crucial that all information is freely available so it may be scrutinised and replicated. This includes all geometric and material parameters, the theoretical formulation and numerical implementation of the simulation, and the full suite of calculation results. While we cannot include all of this data in this paper, this Section contains a selection of our suggested benchmark studies, including their defining parameters and the solutions as calculated with \numbat. Further reference examples and full data are contained in \numbat's online repository~\cite{numbatcode}.

Here we present four reference examples that cover backward SBS, forward intramodal SBS and forward intermodal SBS, in various waveguide geometries and materials that are included in the \numbat suite of literature examples. Each reference case has reliable published 
experimental results:
\begin{itemize}
\item Forward intramodal SBS in a rectangular silicon waveguide, as studied by Van Laer {\em et al.}~\cite{VanLaer:NP:2015}
\item Backward SBS self-cancellation in a silica nanowire, as studied by Florez {\em et al.}~\cite{Florez:NC:2016}
\item Forward SBS in a silicon rib waveguide, as studied by Kittlaus {\em et al.}~\cite{Kittlaus:2017}
\item Backward SBS in a chalcogenide rib waveguide, as studied by Morrison {\em et al.}~\cite{Morrison:Optica:2017}
\end{itemize}
Note that a convergence study for one problem is included in Appendix A.

\subsection{Benchmark 1: Forward intramodal SBS in a rectangular silicon waveguide, (NumBAT literature example 3.6.4)}  
\label{FSBS_sect}

\begin{figure}
\begin{center}
   \includegraphics[width=0.3\textwidth]{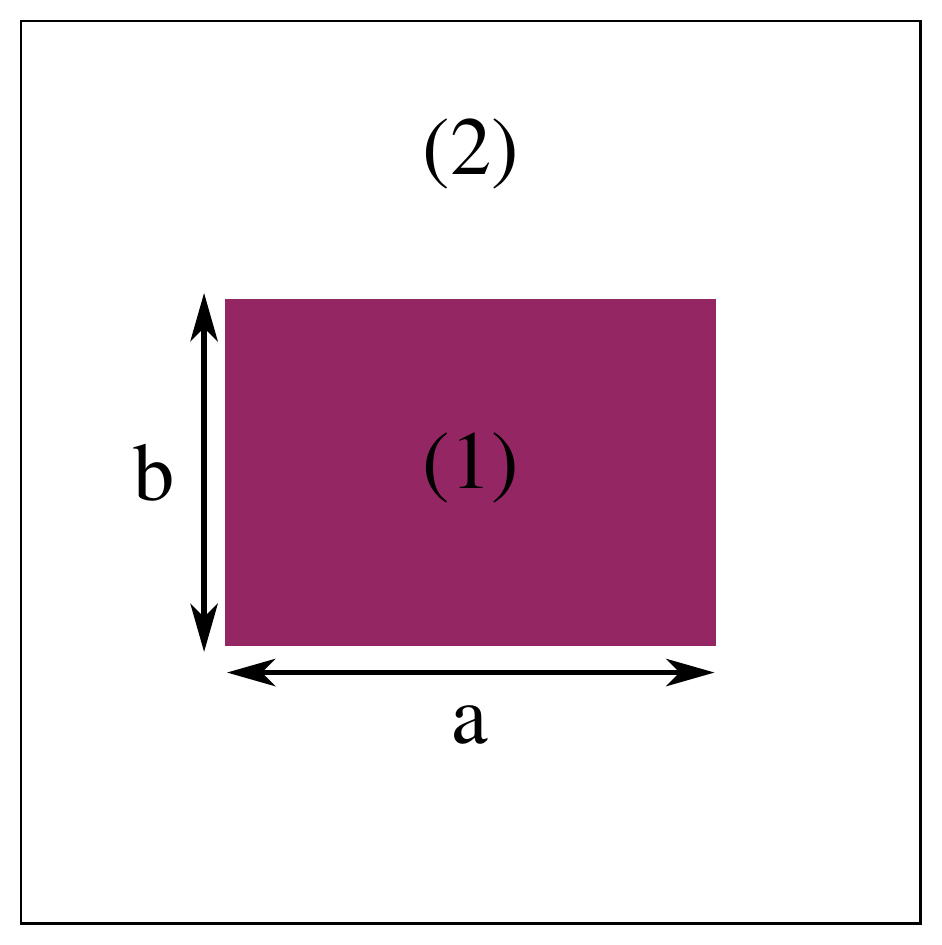}
\end{center}
\caption{Cross-section of a rectangular silicon waveguide (1) surrounded by air (2).}
\label{fig:rectangle:ex1:Tutorial}
\end{figure}

We chose to replicate the results in~\cite{VanLaer:NP:2015} as an experimental demonstration of forward intramodal SBS. While the fabricated sample of~\cite{VanLaer:NP:2015} has the rectangular silicon waveguide supported by a nanoscale pedestal, here we make the simplification that it is suspended in air as illustrated in Fig.~\ref{fig:rectangle:ex1:Tutorial}. This makes for a better canonical reference case and brings the example in line with the original structures proposed for ``giant SBS'' in nanophotonic 
waveguides~\cite{Rakich:PRX:2012}. 
In our calculation, the waveguide has a rectangular cross-section (see Fig.~\ref{fig:rectangle:ex1:Tutorial}), with side lengths $a = 485 \, {\rm nm}$ and $b = 230 \, {\rm nm}$. Note that we have increased the width by $10\%$ from the values in~\cite{VanLaer:NP:2015} to accommodate for the bulging shape 
of the fabricated sample.

\numbat's online repository~\cite{numbatcode} contains a simulation of the fields in the presence of the pedestal; the elastic leakage through the pedestal is however not yet handled in the current version of \numbat, which will require the addition of a surrounding elastic
perfectly matched layer.

For silicon we use the material properties from~\cite{Smith:Wolff:OL:2016}: a refractive index of 3.48; a density of 2329~kg/m$^3$; stiffness tensor components of $c_{11} = 165.6 \, {\rm GPa}$, $c_{12} = 63.9 \, {\rm GPa}$, $c_{44} = 79.5 \, {\rm GPa}$; photoelastic tensor components of $p_{11} = -0.094$, $p_{12} = 0.017$, $p_{44} = -0.051$; and elastic loss tensor components of 
$\eta_{11} = 0.0059 \, \mbox{Pa}\cdot\mbox{s}$, $\eta_{12} = 0.00516 \, \mbox{Pa}\cdot\mbox{s}$, $\eta_{44} = 0.00062 \, \mbox{Pa}\cdot\mbox{s}$. These reference values are for silicon with $\langle 100\rangle$ crystalline orientation, whereas the experiment is for $\langle 110\rangle$ silicon, so we use \numbat's inbuilt function {\ttfamily rotate\_tensor} to rotate the (fourth-rank) tensors appropriately \cite{Auld:book:1973}.

\begin{figure}
\begin{center}
   \includegraphics[width=0.5\textwidth]{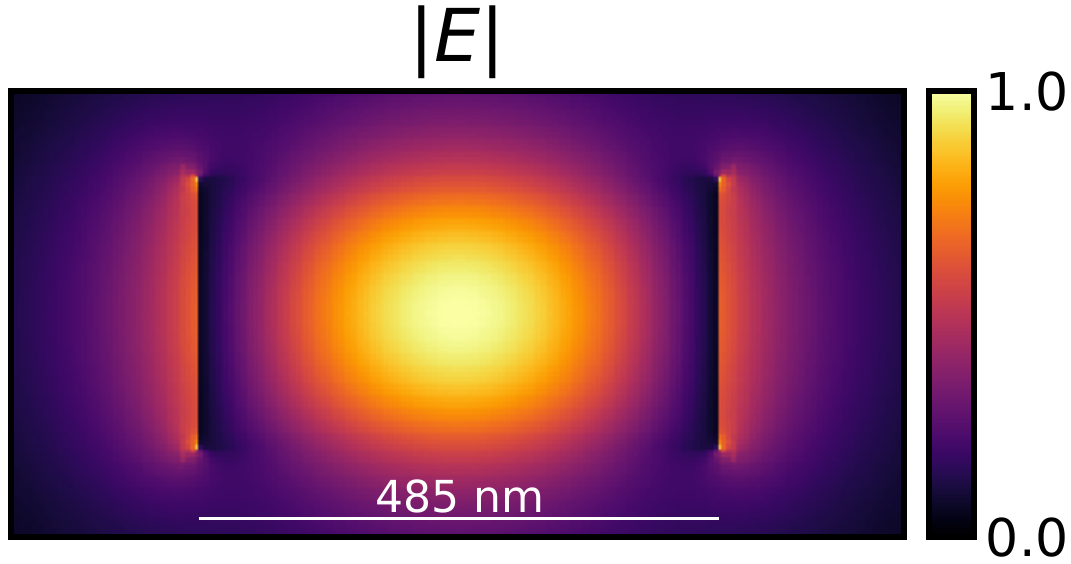}
\end{center}
\caption{Fundamental optical mode profile for the problem in Benchmark 1.}
\label{fig:VanLaer-NP-EM_E_field_0}
\end{figure}

\begin{figure}
\begin{center}
   \includegraphics[width=0.5\textwidth]{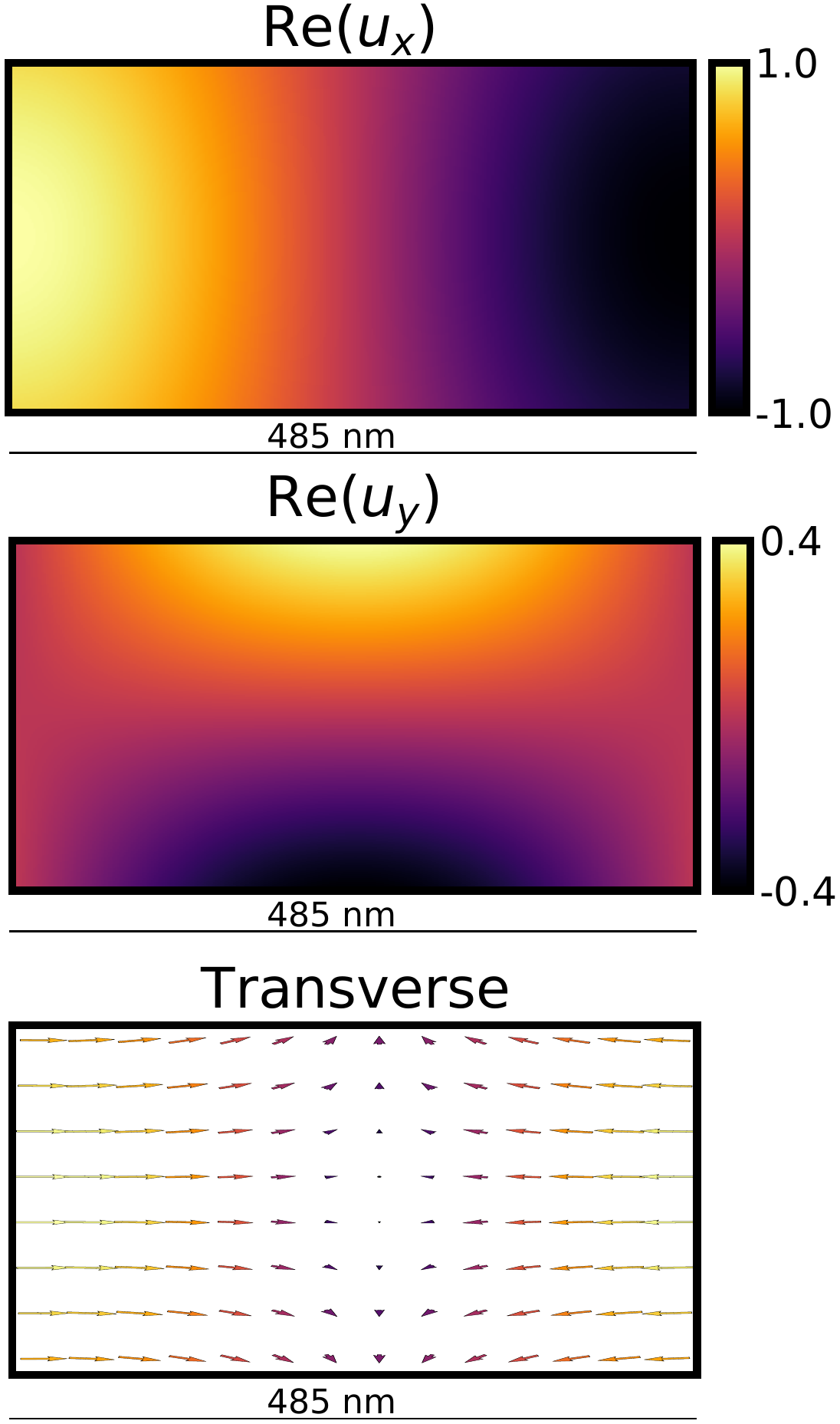}
\end{center}
\caption{Field components of the elastic mode at 9.2~GHz for the problem in Benchmark~1.}
\label{fig:VanLaer-NP-AC_field_6}
\end{figure}

The field profile of the fundamental optical mode, at the wavelength $\lambda = 1550 \, {\rm nm}$, is shown in Fig.~\ref{fig:VanLaer-NP-EM_E_field_0}, and the elastic mode profile is shown in Fig.~\ref{fig:VanLaer-NP-AC_field_6}. Note that \numbat ignores air regions for elastic calculations, \emph{i.e.} treating the air as vacuum, so elastic mode plots (eg. Fig.~\ref{fig:VanLaer-NP-AC_field_6}) only show results in regions of non-air materials.

\begin{figure}
\begin{center}
   \includegraphics[width=0.5\textwidth]{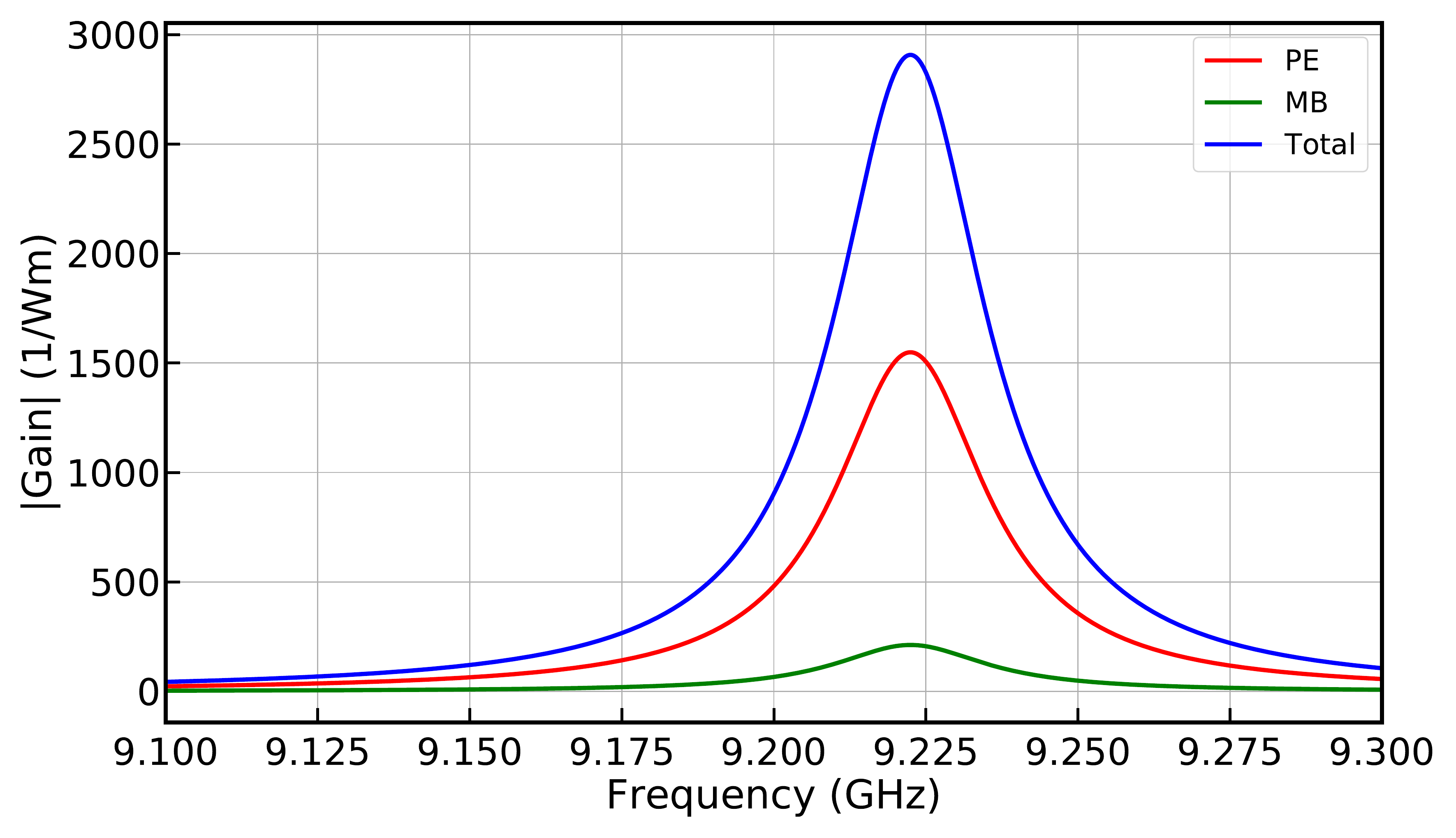}
\end{center}
\caption{Gain spectrum for the problem in Benchmark~1, equivalent to Fig.~2a in~\cite{VanLaer:NP:2015}. The minor difference in centre frequency compared with the paper is due to the non-square shape of the fabricated sample as well as the supporting pedestal.}
\label{fig:lit_04-no_pillar-gain_spectra-MB_PE_comps}
\end{figure}

The forward SBS gain shown in Fig.~\ref{fig:lit_04-no_pillar-gain_spectra-MB_PE_comps} is in good agreement with the experimental result of Fig.~2a of~\cite{VanLaer:NP:2015}, with a clear peak around 9.2~GHz.
In Fig.~\ref{fig:lit_04-no_pillar-gain_spectra-MB_PE_comps}, the legends ``PE'', ``MB'' and ``Total'' correspond respectively to the gain due to the photoelastic effect $Q^{\rm PE}$ (see Eq.~(\ref{PE})) alone, the gain due to moving boundary effect $Q^{\rm MB}$ (see Eq.~(\ref{MB})) alone, and the total gain $Q^{\rm Tot}$ (see Eq.~(\ref{Q:Tot})).
The peak gain in the spectrum is 2907~$\mbox{W}^{-1}\cdot\mbox{m}^{-1}$, which is close to the experimental value of 3200~$\mbox{W}^{-1}\cdot\mbox{m}^{-1}$. The minor discrepancy in frequency and gain is due to the non-square shape of the fabricated sample, the supporting pedestal, as well as minor differences in material properties \footnote{See supplementary materials of ~\cite{VanLaer:NP:2015}.}.
Observe that the peak values satisfy the relation $\Gamma^{\text{Total}}=(\sqrt{\Gamma^\text{PE}} + \sqrt{\Gamma^\text{MB}})^2$.

\subsection{Benchmark 2: SBS self-cancellation in a silica nanowire, (NumBAT literature example 3.6.6)}  

It is possible to design waveguides such that the gain components $Q^{\rm PE}$ and $Q^{\rm MB}$ satisfy $Q^{\rm PE} = -Q^{\rm MB}$, which causes the cancellation of the total gain $Q^{\rm Tot} = Q^{\rm PE} + Q^{\rm MB}$.
This phenomenon provides an excellent numerical test because any error in the sign or magnitude of the radiation pressure or moving boundary terms will be readily apparent in the absence of the cancellation.
This Brillouin scattering self-cancellation effect was first demonstrated by Florez {\em et al.}~\cite{Florez:NC:2016}, whose structure we replicate here.

\begin{figure}
\begin{center}
   \includegraphics[width=0.3\textwidth]{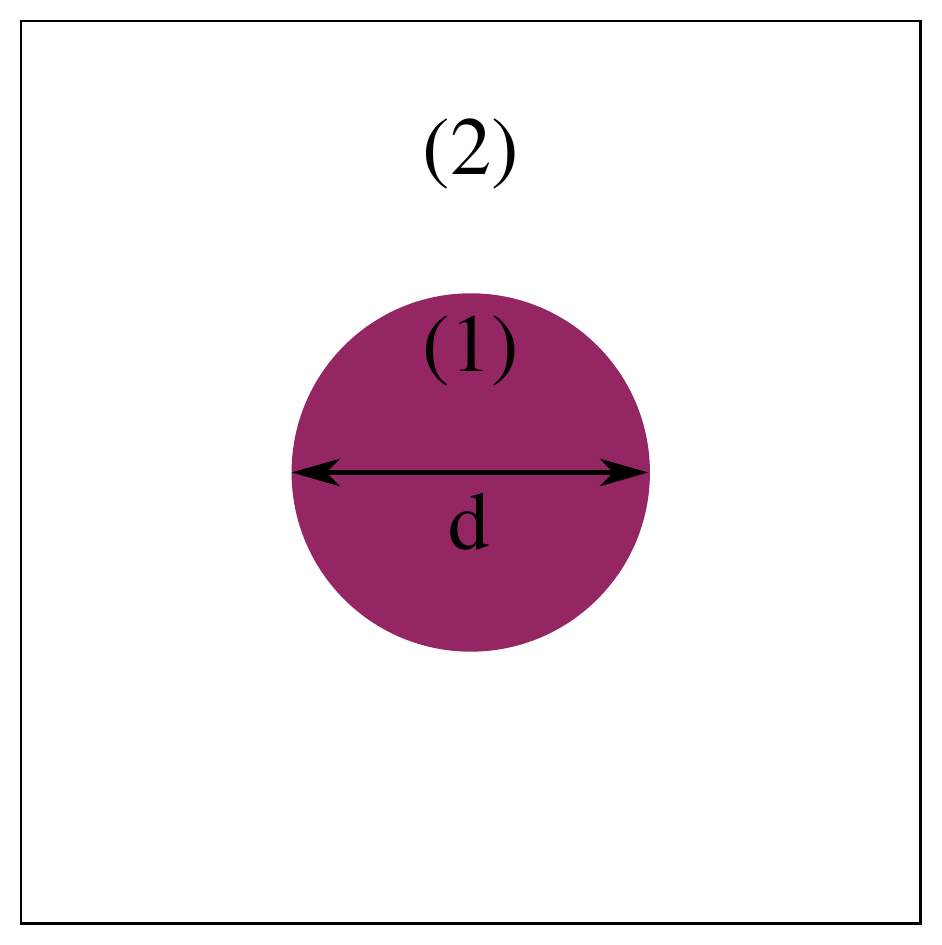}
\end{center}
\caption{Cross-section of a circular wire (1) surrounded by air (2).}
\label{fig:circular:wire:ex2:Florez}
\end{figure}

As illustrated in Fig.~\ref{fig:circular:wire:ex2:Florez}, the waveguide is a silica nanowire with a circular cross-section of diameter $d$ (refractive index $n_1$), surrounded by air (refractive index $n_2$). 
For silica we use the material properties from~\cite{Laude:AIP:2013}: a refractive index of 1.44; a density of 2203~kg/m$^3$; stiffness tensor components of $c_{11} = 78 \, {\rm GPa}$, $c_{12} = 16 \, {\rm GPa}$, $c_{44} = 31 \, {\rm GPa}$; photoelastic tensor components of $p_{11} = 0.12$, $p_{12} = 0.270$, $p_{44} = -0.073$; and elastic loss tensor components of $\eta_{11} = 0.0016 \, \mbox{Pa}\cdot\mbox{s}$, $\eta_{12} = 0.00129 \, \mbox{Pa}\cdot\mbox{s}$, $\eta_{44} = 0.00016 \, \mbox{Pa}\cdot\mbox{s}$.
The wavelength of the incident optical wave is $\lambda = 1550~{\rm nm}$.

\begin{figure}
\begin{center}
   \includegraphics[width=0.5\textwidth]{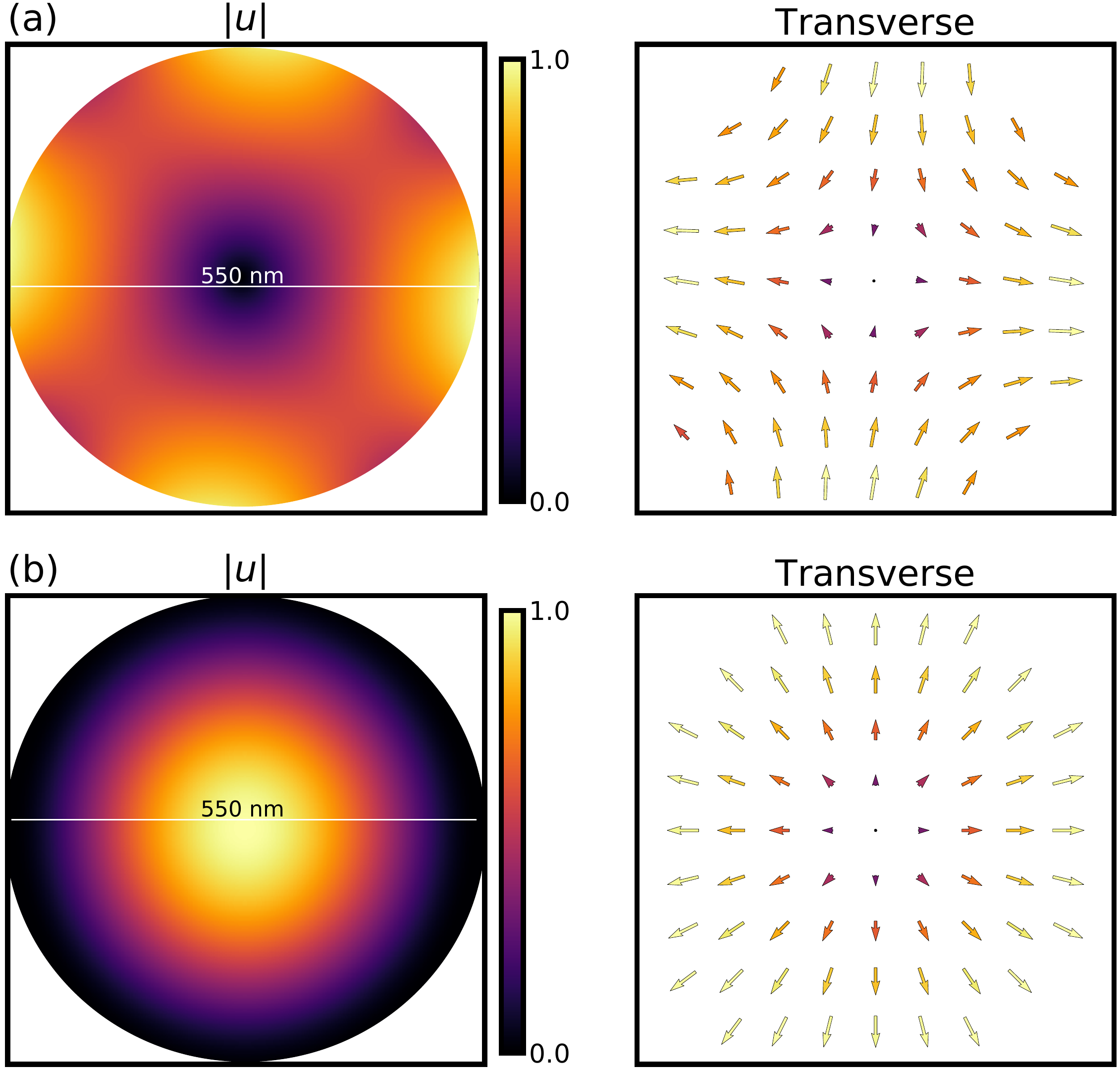}
\end{center}
\caption{Mode profiles of the two lowest order elastic modes of a silica nanowire with a diameter $d = 550 \, {\rm nm}$ for Benchmark 2, (see also Fig.~2(a,b) in Florez {\em et al.}~\cite{Florez:NC:2016}).
The elastic frequency of the modes in (a) and (b) are respectively 5.88~GHz and 6.30~GHz.
}
\label{fig:AC:field:Florez}
\end{figure}

For a nanowire of diameter $d = 550 \, {\rm nm}$, the magnitude $|\bu|$ and the transverse displacement field $(u_x,u_y)$,
of the two lowest order elastic modes, are plotted in Fig.~\ref{fig:AC:field:Florez}(a) and (b).
The transverse components of the modes in the panels~(a) and (b) of Fig.~\ref{fig:AC:field:Florez} have respectively an axially asymmetric torsional-radial profile ($\mbox{TR}_{21}$~mode) and an axially symmetric radial profile ($\mbox{R}_{01}$~mode).

In Fig.~\ref{fig:AC:field:Florez}(b), although the transverse displacement field of the $R_{01}$~mode has small values near the waveguide centre, we observe that  the total magnitude $|\bu|$ of the $\mbox{R}_{01}$~mode is large near the centre. This is due to the fact that the longitudinal displacement $u_z$ takes large values near the waveguide centre, which is consistent with the curves in Fig.~5a of~\cite{Florez:NC:2016}.

\begin{figure}
\begin{center}
   \includegraphics[width=0.5\textwidth]{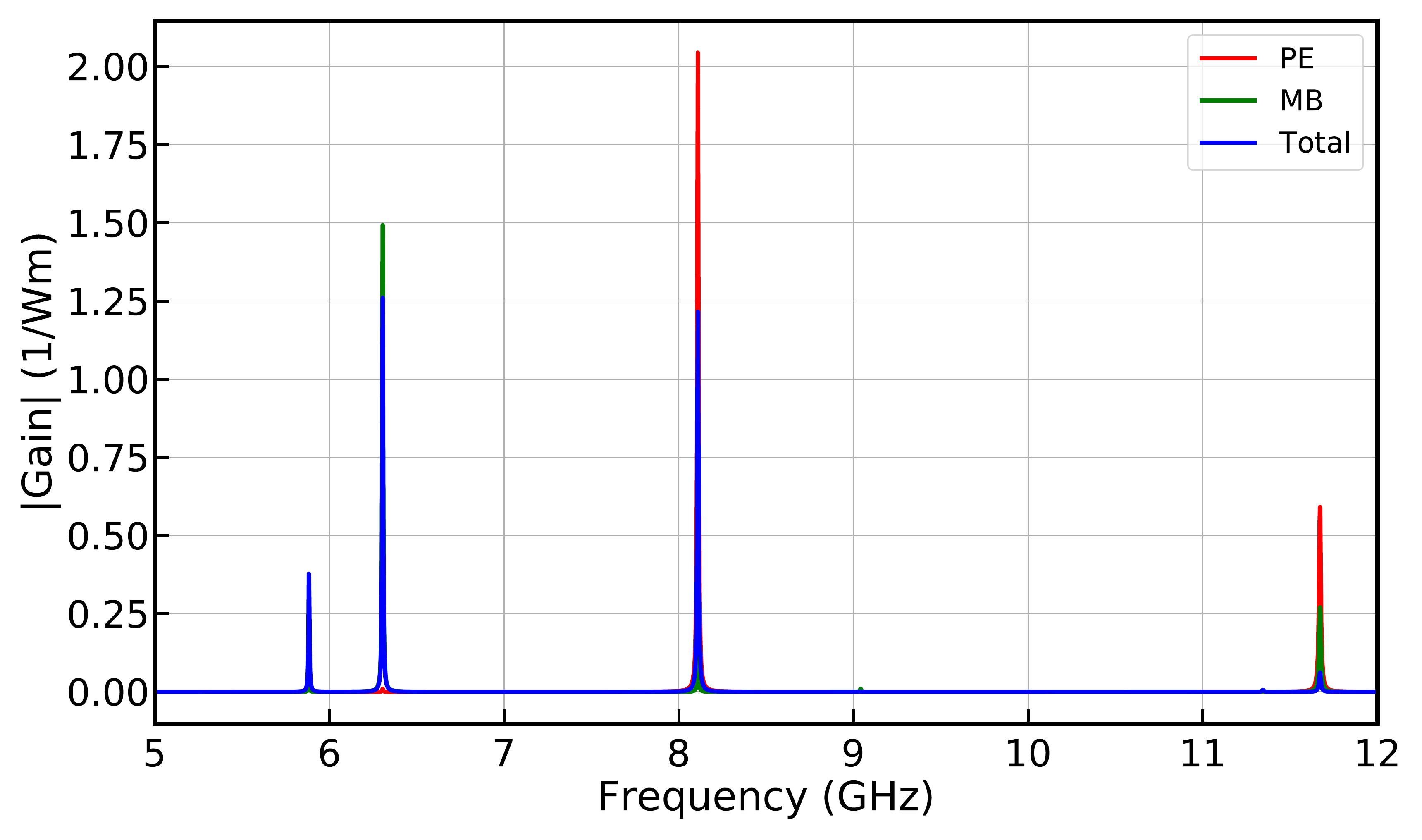}
\end{center}
\caption{Gain spectra of a silica nanowire with diameter 550 nm for Benchmark 2, matching the blue curve of Fig.~3b in Florez {\em et al.}~\cite{Florez:NC:2016}.}
\label{fig:gain:spectra:MB:PE:Florez}
\end{figure}

\begin{figure}
\begin{center}
   \includegraphics[width=0.5\textwidth]{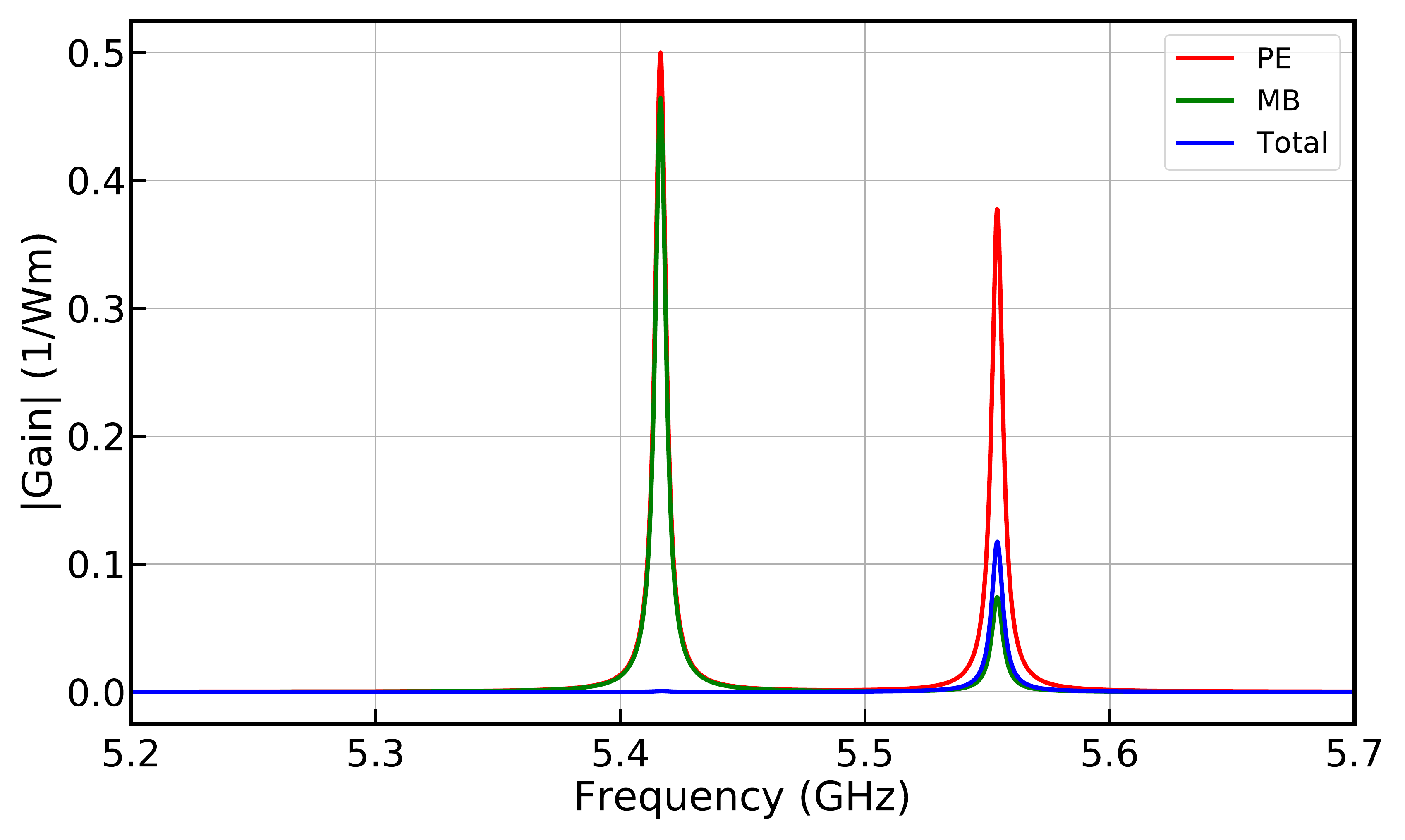}
\end{center}
\caption{Gain spectra of a silica nanowire with diameter $d=1160\, {\rm nm}$ for Benchmark 2, as in Fig.~4 of Florez {\em et al.}~\cite{Florez:NC:2016}, showing near perfect cancellation at 5.4 GHz.}
\label{fig:gain:spectra:MB:PE:Florez:2}
\end{figure}

\begin{figure}
\begin{center}
   \includegraphics[width=0.5\textwidth]{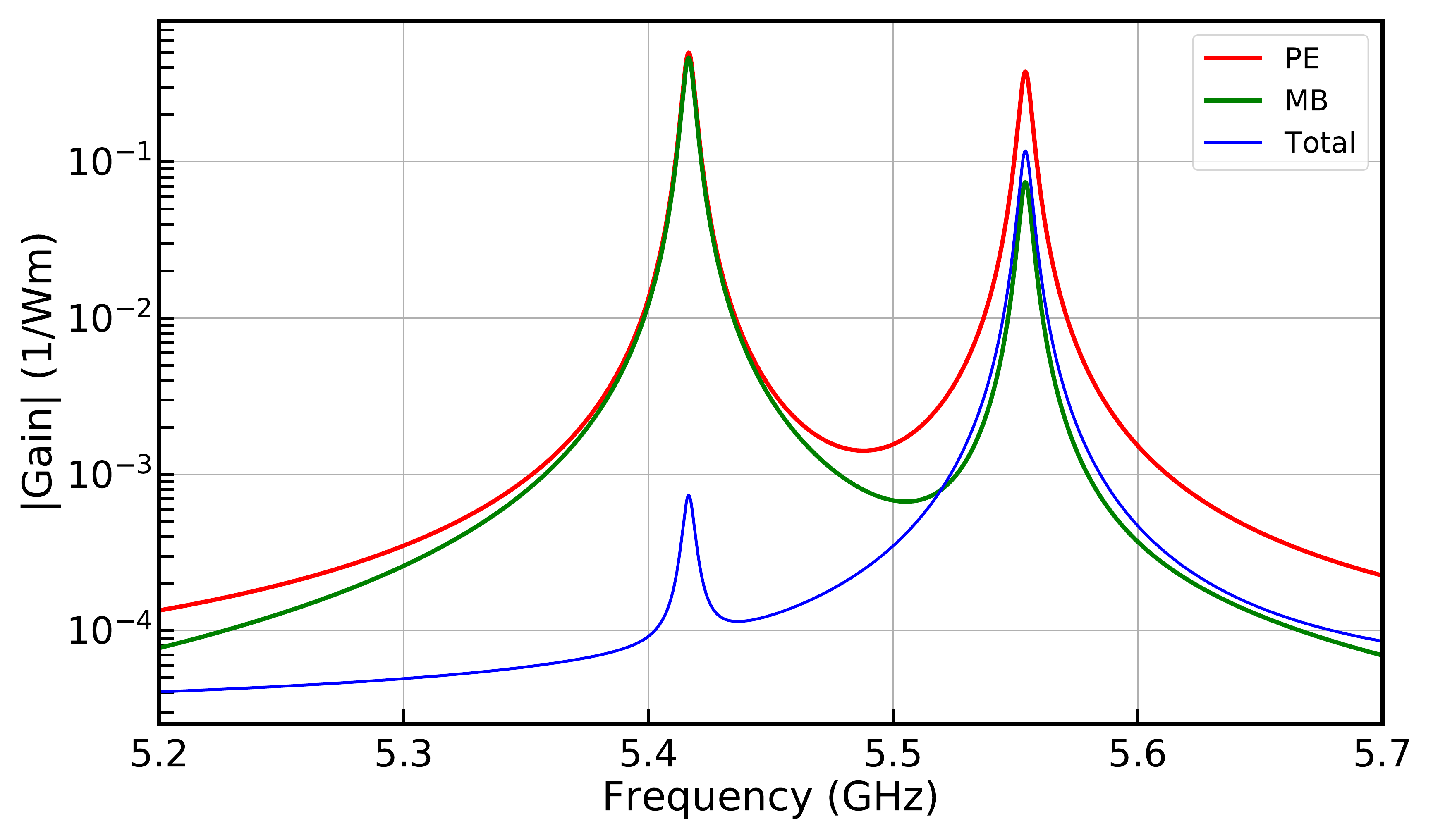}
\end{center}
\caption{The same gain spectra as in Fig.~\ref{fig:gain:spectra:MB:PE:Florez:2} plotted on a log-scale.}
\label{fig:gain:spectra:MB:PE:logy:Florez}
\end{figure}

The total gain spectrum of the silica nanowire with a diameter 550~nm is shown in blue in Fig.~\ref{fig:gain:spectra:MB:PE:Florez} and matches the blue curve of Fig.~3b in~\cite{Florez:NC:2016}.
Both the $\mbox{TR}_{21}$~mode (elastic frequency 5.88~GHz) and the $\mbox{R}_{01}$~mode (elastic frequency 6.30~GHz) display a significant total gain in Fig.~\ref{fig:gain:spectra:MB:PE:Florez}. As observed in~\cite{Florez:NC:2016}, the total gain of the $\mbox{TR}_{21}$~mode is dominated by the contribution from the photoelastic effect, while both the photoelastic effect and the moving boundary effect can play a significant role in the total gain of the $\mbox{R}_{01}$~mode.
In particular, when the diameter of the nanowire is $d = 1160\,{\rm nm}$, for the $\mbox{R}_{01}$~mode, the contributions from the photoelastic effect and moving boundary effect are equal and opposite, resulting in a Brillouin scattering self-cancellation.
This is confirmed by the gain spectrum in Fig.~\ref{fig:gain:spectra:MB:PE:Florez:2}, which shows near perfect cancellation at 5.4~GHz.
For clarity, the curves in Fig.~\ref{fig:gain:spectra:MB:PE:Florez:2} are shown on a log-scale in Fig.~\ref{fig:gain:spectra:MB:PE:logy:Florez}, revealing that the cancellation holds to approximately one part in a thousand.
The gain spectra in Figs.~\ref{fig:gain:spectra:MB:PE:Florez:2} and \ref{fig:gain:spectra:MB:PE:logy:Florez} are in agreement with 
the results in Fig.~4 of~\cite{Florez:NC:2016}.

%
\subsection{Benchmark 3: Forward intermodal SBS in a silicon rib waveguide, (NumBAT literature example 3.6.8)}  
Next we simulate forward intermodal SBS in a silicon rib waveguide. This structure has been used in recent experiments by Kittlaus {\em et al.}~\cite{Kittlaus:2017} and consists of a ridge that is 80~nm high and 1500~nm wide upon a membrane that is 135~nm and 2850~nm wide. Note that Fig.~2e in~\cite{Kittlaus:2017} shows the height of the membrane to be 145~nm while the text describes the layer as being 135~nm thick; we use 135~nm here. Once again the crystalline orientation of silicon is $\langle 110 \rangle$ and the material properties are the same as in Sect.~\ref{FSBS_sect}.

\begin{figure}
\begin{center}
   \includegraphics[width=0.5\textwidth]{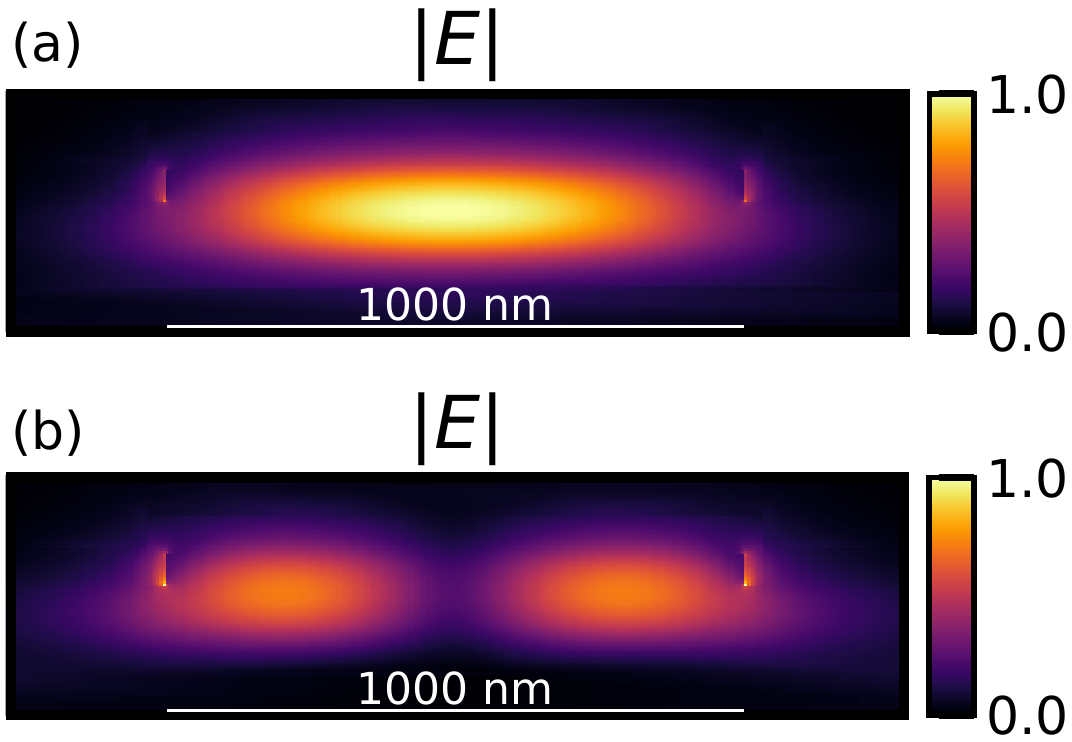}
\end{center}
\caption{Fundamental (a) and second order (b) optical mode fields of the Benchmark 3 structure of~\cite{Kittlaus:2017}, equivalent to Fig.~2f,g in~\cite{Kittlaus:2017}.}
\label{fig:EM:E:field:0:Kittlaus}
\end{figure}

\begin{figure}
\begin{center}
   \includegraphics[width=0.5\textwidth]{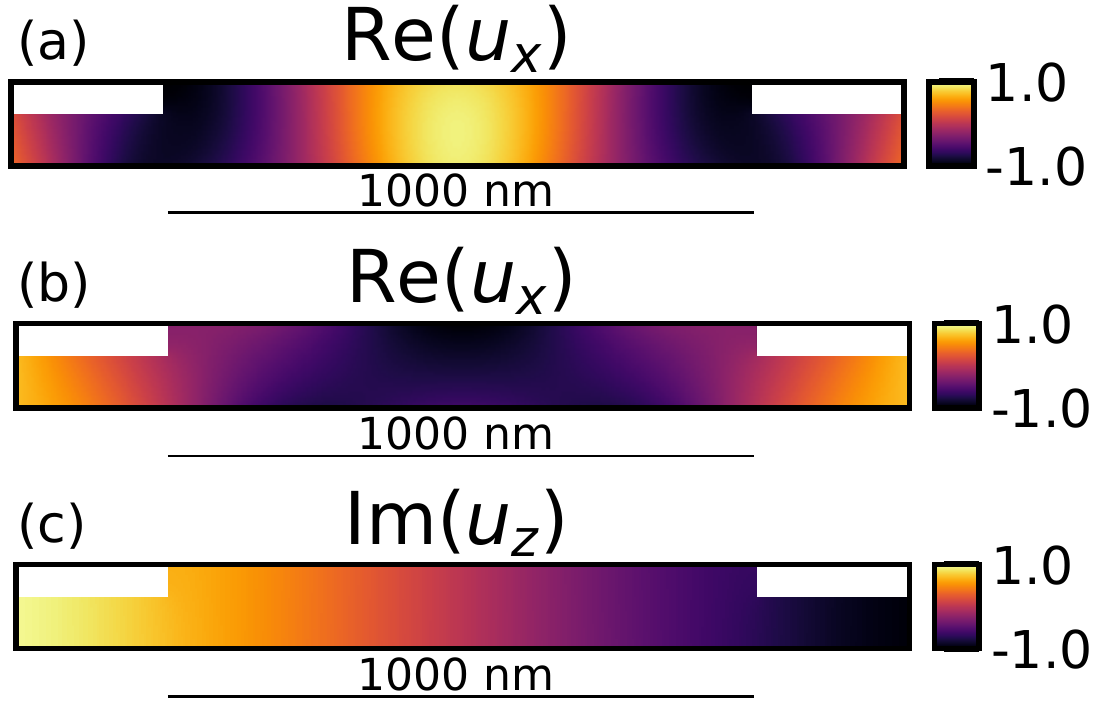}
\end{center}
\caption{Dominant high gain elastic modes for Benchmark 3 as shown in Fig.~3c of~\cite{Kittlaus:2017}.
The elastic frequency of the modes in (a), (b) and (c) are respectively 5.93~GHz, 3.03~GHz, and 1.33~GHz.}
\label{fig:AC:field:Kittlaus}
\end{figure}

Figure~\ref{fig:EM:E:field:0:Kittlaus} shows the fundamental optical mode fields of the structure, equivalent to Fig.~2f,g in~\cite{Kittlaus:2017}. Figure~\ref{fig:AC:field:Kittlaus} shows the elastic mode fields of the modes at 5.93~GHz, 3.03~GHz, and 1.33~GHz as shown in Fig.~3c of~\cite{Kittlaus:2017}.

\begin{figure}
\begin{center}
   \includegraphics[width=0.5\textwidth]{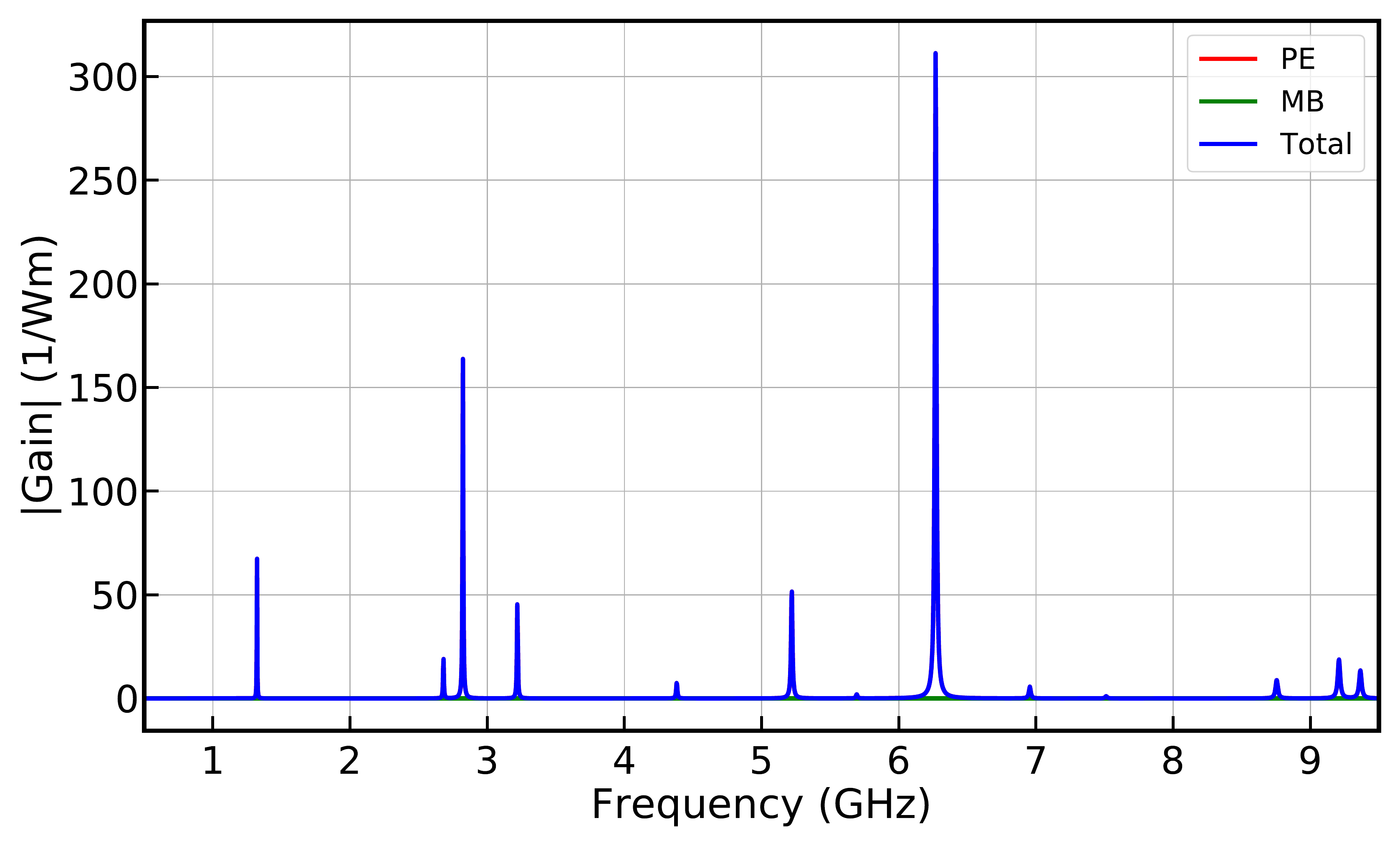}
\end{center}
\caption{Gain spectra for Benchmark 3 equivalent to Fig.~3b in~\cite{Kittlaus:2017}.}
\label{fig:gain:spectra:MB:PE:Kittlaus}
\end{figure}

Figure~\ref{fig:gain:spectra:MB:PE:Kittlaus} shows the gain spectra equivalent to Fig.~3b in~\cite{Kittlaus:2017}. Note that the gain is completely dominated by the photoelastic effect---this is because the only non-cancelling contribution 
from the moving boundaries comes from the small vertical sections at the side of the central rib, 
where both the optical and elastic intensities are extremely small. 

\subsection{Benchmark 4: Backward SBS in a chalcogenide rib waveguide, (NumBAT literature example 3.6.10)}

Stimulated Brillouin scattering in integrated devices was first demonstrated in partially etched rib waveguides made of a soft amorphous glass, $\mbox{As}_2\mbox{S}_3$~\cite{Eggleton2013}. The large photoelastic constants make strong Brillouin interactions possible in even large mode area waveguides, with large amplification factors beyond 50~dB being generated in certain structures~\cite{Choudhary:2017}. Here we perform simulations of a recent experimental design from Morrison {\em et al.}: a fully etched  $\mbox{As}_2\mbox{S}_3$ waveguide clad in $\mbox{SiO}_2$, formed within a wider Si circuit~\cite{Morrison:Optica:2017}. This result is a useful benchmark due to its distinctive frequency spectrum, comprising of multiple elastic modes with varying opto-elastic overlaps. Unlike the earlier examples, this particular waveguide geometry is highly multimoded for the elastic wave, with over  100 elastic modes 
needing to be accounted for when finding the SBS gain.

\begin{figure}
\begin{center}
   \includegraphics[width=0.5\textwidth]{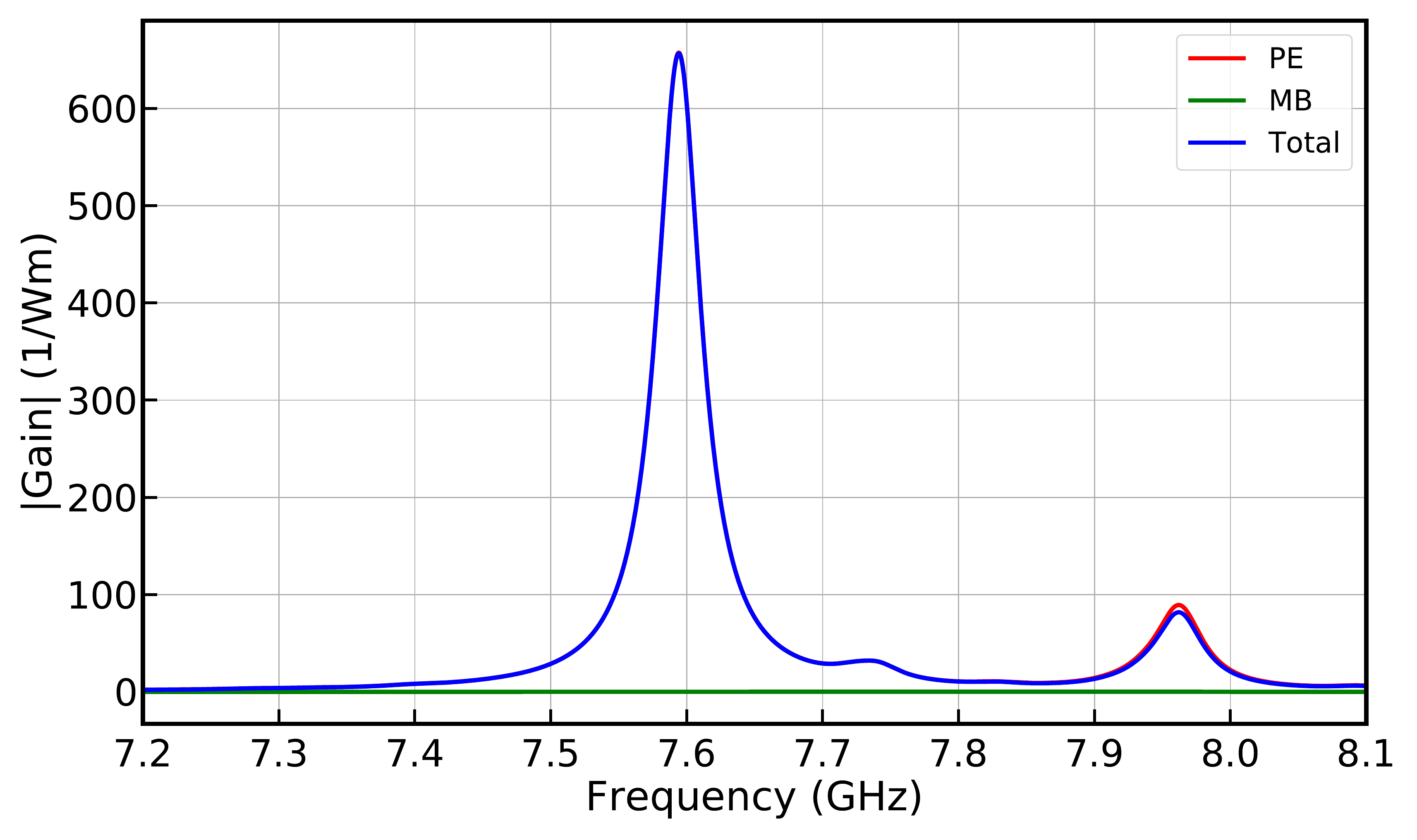}
\end{center}
\caption{Gain spectra for Benchmark 4 equivalent to Fig.~3c in~\cite{Morrison:Optica:2017}.}
\label{fig:gain:spectra:MB:PE:Morrison}
\end{figure}

Figure \ref{fig:gain:spectra:MB:PE:Morrison} shows the gain spectra equivalent to Fig.~3c in~\cite{Morrison:Optica:2017} and a peak gain of 660~$\mbox{W}^{-1}\cdot\mbox{m}^{-1}$, which is almost exclusively driven by the photoelastic effect due to the large cross section area of the waveguide. For $\mbox{As}_2\mbox{S}_3$ we use a refractive index of 2.44; a density of 3150~kg/m$^3$; stiffness tensor components of $c_{11} = 19.75 \, {\rm GPa}$, $c_{12} = 8.7 \, {\rm GPa}$, $c_{44} = 5.525 \, {\rm GPa}$; photoelastic tensor components of $p_{11} = 0.25$, $p_{12} = 0.23$, $p_{44} = 0.01$; and elastic loss tensor components of $\eta_{11} = 0.0018 \, \mbox{Pa}\cdot\mbox{s}$, $\eta_{12} = 0.00145 \, \mbox{Pa}\cdot\mbox{s}$, $\eta_{44} = 0.00018 \, \mbox{Pa}\cdot\mbox{s}$.

The experimentally measured gain from the example in Fig.~\ref{fig:gain:spectra:MB:PE:Morrison} is 
$(750 \pm 50)\, \mbox{W}^{-1} \cdot \mbox{m}^{-1}$, with the error attributed to uncertainties in the elastic loss tensor components. 

\section{Conclusion}
Standardised tests are an essential component for the development of scientific software. We believe 
\numbat will serve the needs of the optomechanics community by providing such a suite of tests, together with 
code that can be used both for investigations of the underlying physics of SBS as 
well as for the design of new Brillouin-active waveguide systems.

Several extensions of \numbat are envisaged and are in the process of implementation. The first is the 
inclusion of open boundaries for the elastic problem---this is important for a large number of experiments, 
in which the acoustic mode is not strictly confined but can radiate away from the waveguide core through the substrate. 
This process results in an additional component to elastic loss that dominates material losses. 
Such elastic radiation loss can be computed either perturbatively, from the magnitude of the elastic field components
at some distance from the waveguide, or directly, through the imposition of an elastic perfectly-matched
layer on the boundary of the computational domain. A second extension to the \numbat package 
is the inclusion of roto-optic effects,
which manifest in optically-anisotropic materials and structures, such as stacks of thin layers which 
differ in their elastic properties. Recent analysis~\cite{Smith2017} has shown that the Brillouin gain 
resulting from such structures can be significant, even dominating the photo-elastic-induced gain
which is included here. It is a straightforward process to include this contribution within the 
framework of \numbat.

A final extension to \numbat is to the computation of gain in waveguides that are not longitudinally invariant, 
such as photonic crystal waveguides and rib-waveguides with periodic supporting struts. 
The computation of SBS gain in these structures requires an extension of the FEM routine to three-dimensional mode computations, taking into account the changed requirements of phase matching between modes which become more complicated
in periodically-replicated media.

\section*{Acknowledgments}
This work was supported by the Australian Research Council under Discovery Grant DP1601016901.
We thank Christian Wolff for many helpful discussions and provision of comparison data, 
and Eric Kittlaus for useful discussions.

\appendix

\section{Convergence}
\label{convergence}

As an example of \numbat's convergence as a function of the resolution of the FEM mesh we study backward SBS in a rectangular silicon waveguide surrounded by vacuum. This simulation is included in the tutorial~\cite{numbatcode} as it is important to carry out convergence testing of this sort (as well as convergence against the number of modes included) whenever simulating new structures.

\begin{figure}
\begin{center}
   \includegraphics[width=0.5\textwidth]{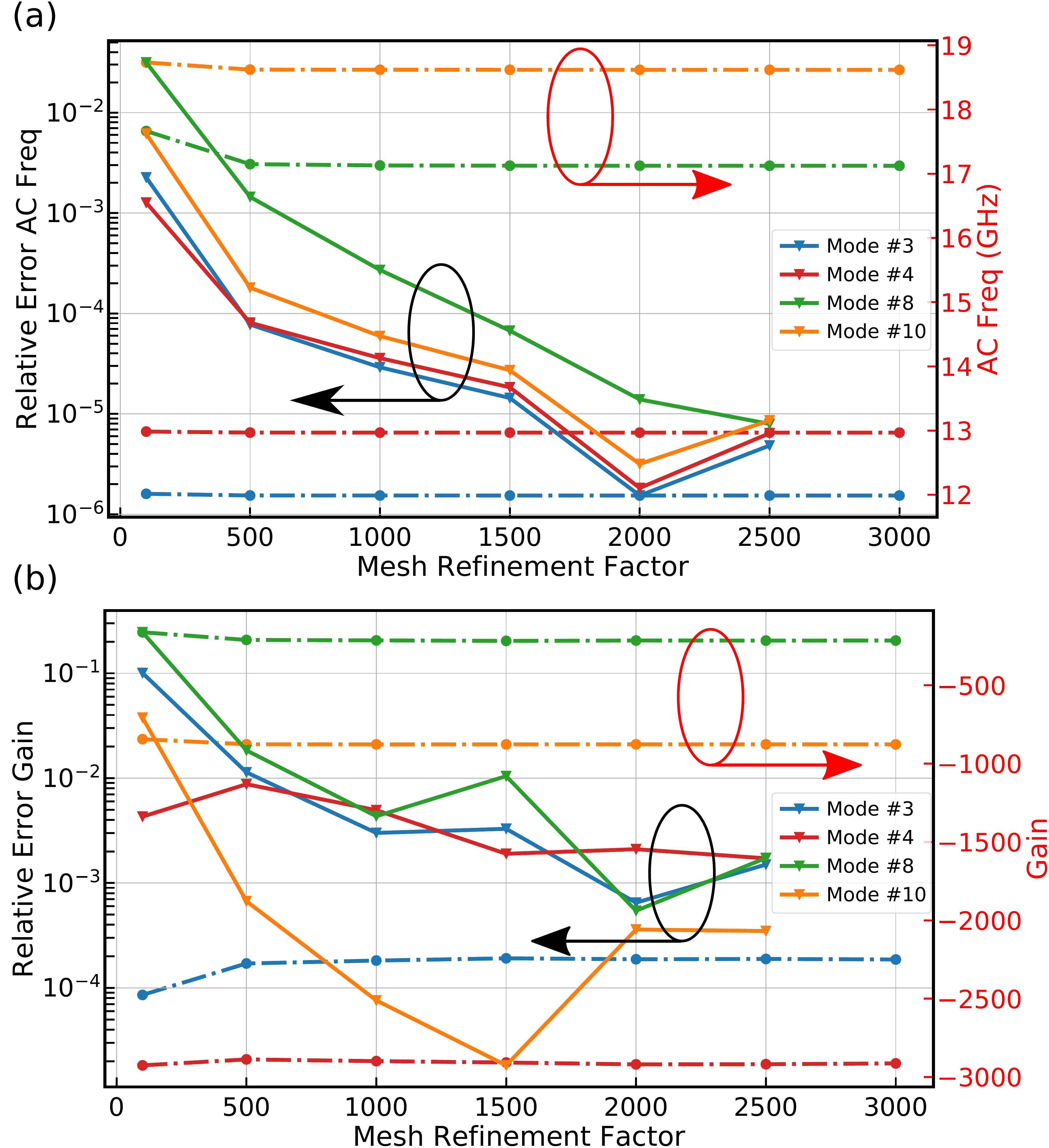}
\end{center}
\caption{Convergence of elastic mode frequency and SBS peak gain in rectangular silicon waveguide.}
\label{fig:conv}
\end{figure}

Figure~\ref{fig:conv}(a) shows the convergence of the elastic frequency, and Fig.~\ref{fig:conv}(b) shows the convergence of the SBS gain, both as a function of the mesh refinement parameter {\ttfamily lc\_2}. This parameter sets the characteristic length of the FEM elements on the boundary of the waveguide as {\ttfamily lc\_bkg / lc\_2}, where {\ttfamily lc\_bkg} is the background mesh size. The larger {\ttfamily lc\_2} the finer the mesh on the waveguide boundary. In each figure we show only those modes that experience significant gain, representing the absolute value of the quantity in dashed lines with circular markers, and the error relative to the most finely resolved mesh in solid lines with triangular markers. 

%
%
%

\section{Finite element calculation of the elastic waveguide modes}
\label{sec:FEM}

\subsection{Equations of elastic waves}
The complex displacement vector $\bu (\br,t)$ of an elastic field undergoing harmonic oscillation takes the form:
\begin{eqnarray}
\bu (\br,t) =
\bu (x,y,z)  \, e^{ - i \, \Omega \, t}= 
\left[ \begin{array}{c}
u_x (x,y,z) \\ u_y (x,y,z) \\ u_z(x,y,z)
\end{array}\right]   \, 
e^{- i \, \Omega \, t} ,
\end{eqnarray}
where $\br$ is the position vector $\br = (x,y,z)$.
The strain tensor $\bS$ corresponding to the displacement vector $\bu$ is
\begin{eqnarray}
\label{strain:tensor}
\bS & = & \nabla_S \, \bu, \\
 ~  & = & \frac{1}{2} \left( \nabla\bu +  \left( \nabla \bu \right)^T \right) .
\end{eqnarray}
%
%
Hooke's Law states that the stress tensor $\bT$ is linearly proportional to the strain tensor $\bS$~(see Eq.~(3.12), p. 43 in~\cite{Auld:book:1990}):
\begin{eqnarray}
\label{Hooke:eq1}
\bT & = & \bc : \bS ,
\end{eqnarray}
where $\bc$ is the fourth-rank stiffness tensor characterising the material.

\medskip
In the frequency domain, i.e., $\bu( \br, t) = \bu(\br) \, e^{-i \, \Omega \, t}$,
the dynamical equation of elasticity in the absence of an external force is~(See Eq.~(2.18), p. 43 in~\cite{Auld:book:1990})
%
\begin{eqnarray}
\label{wave:eq1}
\nabla \cdot \bT + \Omega^2 \, \rho \, \bu & = & 0 ,
\end{eqnarray}
where $\rho$ is the material density.
In order to develop a finite element method solution, the wave equation Eq.~(\ref{wave:eq1}) is written in weak form (variational formulation).
We first take the product of Eq.~(\ref{wave:eq1}) with the conjugate of a test function $\bm{v}$, and integrate  over the domain $A$ of the elastic field:
\begin{eqnarray}
\label{variational:eq1}
\int_{A}
\bm{v}^* \cdot \left( \nabla \cdot \bT + \Omega^2 \, \rho \, \bu \right) \, \mathrm{d}V = 0 \, .
\end{eqnarray}
Since the  stress tensor $\bT$ is symmetric, by applying the identity 
$\nabla \cdot \left( \bT \, \bm{v} \right) 
= \bm{v} \cdot \left( \nabla \cdot \bT \right) 
+ \bT : \nabla \bm{v}$,
we are led to the following relation:
\begin{eqnarray}
\label{div:formula}
\bm{v}^* \cdot \left( \nabla \cdot \bT \right) 
& = &
\nabla \cdot \left( \bT \, \bm{v}^* \right) 
- \nabla_S \bm{v}^* : \bT  ,
\end{eqnarray}
%
where the operator $:$ denotes the double dot product of two second-rank tensors~\cite{Auld:book:1990}.
In the derivation of Eq.~(\ref{div:formula}), we have used the fact that the double dot product of
the anti-symmetric part of $\nabla \bm{v}$ and tensor $\bT$ is zero, since $\bT$ is symmetric.
By substituting Eq.~(\ref{div:formula})  into Eq.~(\ref{variational:eq1})
and applying the divergence theorem together with Eqs.~(\ref{strain:tensor}) and (\ref{Hooke:eq1}), we obtain the following expression after integration by parts:
\begin{eqnarray}
\label{variational:eq2}
\begin{aligned}
\int_{A}
&\left(- \nabla_S \bm{v}^* : \left( \bc : \nabla_S \, \bu \right) + 
\Omega^2 \, \rho \, \bm{v}^* \cdot \bu\right)  \, \dA \\
&+ \int_{\partial A} \bn \cdot \left( \bT \, \bm{v}^*  \right) \, dS 
= 0 \, ,
\end{aligned}
\end{eqnarray}
where $\partial A$ denotes the boundary of the domain A.
The value of the boundary integral in Eq.~(\ref{variational:eq2}) is zero when 
either of the following boundary conditions are applied to $\partial A$:
the free interface condition $\bT \cdot \bn = \bm{0}$
or
the fixed interface condition $\bu = \bm{0}$.
%

%
%

\subsection{Matrix notation}

Since the tensors $\bS$ and $\bT$ are symmetric, they possess only six independent coefficients
and for numerical implementations, it is convenient to represent 
$\bS$ and $\bT$ as six-component vectors (Voigt notation):
\begin{eqnarray}
\label{Voigt:notation}
\bS =
\left[ \begin{array}{c}
S_1 \\ S_2 \\ S_3 \\ S_4 \\ S_5 \\ S_6
\end{array}\right] 
\quad
\mbox{and}
\quad
\bT =
\left[ \begin{array}{c}
T_1 \\ T_2 \\ T_3 \\ T_4 \\ T_5 \\ T_6
\end{array}\right] .
\end{eqnarray}
%
The coefficients $S_1, \dots, S_6$ and $T_1, \dots , T_6$ in Eq.~(\ref{Voigt:notation}) are defined by the relations~\cite{Auld:book:1990}:
\begin{eqnarray}
\bS =
\left[ \begin{array}{ccc}
S_{xx} & S_{xy} & S_{xz} \\[6pt]
S_{xy} & S_{yy} & S_{yz} \\[6pt]
S_{xz} & S_{yz} & S_{zz}
\end{array}\right]
=
\left[ \begin{array}{ccc}
S_{1} & \frac{1}{2} \, S_{6} & \frac{1}{2} \, S_{5} \\[6pt]
\frac{1}{2} \, S_{6} & S_{2} & \frac{1}{2} \, S_{4} \\[6pt]
\frac{1}{2} \, S_{5} & \frac{1}{2} \, S_{4} & S_{3}
\end{array}\right] ,
\end{eqnarray}
and
\begin{eqnarray}
\bT =
\left[ \begin{array}{ccc}
T_{xx} & T_{xy} & T_{xz} \\
T_{xy} & T_{yy} & T_{yz} \\
T_{xz} & T_{yz} & T_{zz}
\end{array}\right]
=
\left[ \begin{array}{ccc}
T_{1} & T_{6} & T_{5} \\
T_{6} & T_{2} & T_{4} \\
T_{5} & T_{4} & T_{3}
\end{array}\right],
\end{eqnarray}
noting the absence of the $\tfrac{1}{2}$ factors in the stress tensor $\bT$.
In this matrix notation, the definition $\bS = \nabla_S \, \bu$ and the wave equation Eq.~(\ref{variational:eq2}) 
still hold if 
the matrix form of the symmetric gradient operator $\nabla_S$ is used:
\begin{eqnarray}
\label{nabla:S:matrix}
\nabla_S \, \bu
=
\left[ \begin{array}{ccc}
\displaystyle
\frac{\partial}{\partial x} & 0 & 0 \\[12pt]
0 & \displaystyle \frac{\partial}{\partial y} & 0 \\[12pt]
0 & 0 & \displaystyle \frac{\partial}{\partial z} \\[12pt]
0 & \displaystyle \frac{\partial }{\partial z} & \displaystyle \frac{\partial}{\partial y} \\[12pt]
\displaystyle \frac{\partial}{\partial z} & 0 & \displaystyle \frac{\partial}{\partial x} \\[12pt]
\displaystyle \frac{\partial}{\partial y} & \displaystyle \frac{\partial}{\partial x} & 0
\end{array}\right] 
\left[ \begin{array}{c}
u_x \\ u_y \\ u_z
\end{array}\right] .
\end{eqnarray}
In the matrix notation, the fourth-rank stiffness tensor $\bm{c}$ becomes a matrix of order $6 \times 6$.
For example, in media of monoclinic symmetry  with one symmetry plane $(xy)$, 
Hooke's Law can be written as
\begin{eqnarray}
\label{C:matrix:monoclinic}
\left[ \begin{array}{c}
T_1 \\ T_2 \\ T_3 \\ T_4 \\ T_5 \\ T_6
\end{array}\right] 
=
\left[ \begin{array}{cccccc}
c_{11} & c_{12} & c_{13} & 0 & 0 & c_{16} \\
c_{21} & c_{22} & c_{23} & 0 & 0 & c_{26} \\
c_{31} & c_{32} & c_{33} & 0 & 0 & c_{36} \\
 0 & 0 & 0 & c_{44} & c_{45} & 0 \\
 0 & 0 & 0 & c_{45} & c_{55} & 0 \\
 c_{61} & c_{62} & c_{63} & 0 & 0 & c_{66}
\end{array}\right] 
\left[ \begin{array}{c}
S_1 \\ S_2 \\ S_3 \\ S_4 \\ S_5 \\ S_6
\end{array}\right] ,
\end{eqnarray}
where the coefficients $c_{IJ}$ are the elastic stiffness constants of the material 
in the Voigt notation~\cite{Auld:book:1973}. In this notation, the two subscripts $I$ and $J$ stand for the four subscripts of the fourth-rank stiffness tensor $c_{ijkl}$ in pairs $ij$ and $kl$ 
according to the mapping: $1\to 11$, $2\to 22$, $3\to 33$, $4\to 23$, $5\to 13$, $6\to 12$.
So for example, $c_{11}\to c_{1111}$, $c_{16}\to c_{1112}$, and $c_{45}\to c_{2313}$.
%

%
%

\subsection{Waveguide modes}
For elastic waveguide modes, the displacement vector $\bu (\br,t)$ takes the form
\begin{eqnarray}
\begin{aligned}
\bu (\br,t) &=
\bu (x,y)  \, e^{i \, ( q \, z - \Omega \, t)}\\
&= \left[ \begin{array}{c}
u_x (x,y) \\ u_y (x,y) \\ u_z(x,y)
\end{array}\right]   \, 
e^{i \, ( q \, z - \Omega \, t)} ,
\end{aligned}
\end{eqnarray}
where $q$ is the propagation constant.
Since the waveguide mode $\bu$ has an exponential $z$-dependence, the symmetric gradient operator in Eq.~(\ref{nabla:S:matrix}) now takes the form:
\begin{eqnarray}
\label{nabla:S:mode:eq1}
\nabla_S \, \bu
=
\left[ \begin{array}{ccc}
\displaystyle
\frac{\partial }{\partial x} & 0 & 0 \\[12pt]
\displaystyle
0 & \displaystyle \frac{\partial }{\partial y} & 0\\[12pt]
\displaystyle
0 & 0 & \displaystyle i \, q \\[12pt]
\displaystyle
0 & \displaystyle i \, q & \displaystyle \frac{\partial }{\partial y} \\[12pt]
\displaystyle
i \, q & 0 & \displaystyle \frac{\partial }{\partial x} \\[12pt]
\displaystyle
\frac{\partial }{\partial y} & \displaystyle \frac{\partial }{\partial x} & 0
\end{array}\right]
\,
\left[ \begin{array}{c}
u_x \\[12pt]  u_y \\[12pt]  u_z
\end{array}\right] .
\end{eqnarray}
For the problems considered in this work, the propagation constant $q$ has a fixed value and the associated frequency $\Omega$
is the unknown eigenvalue.
By imposing the free interface condition $\bT \cdot \bn = \bm{0}$
(or the fixed interface condition $\bu = \bm{0}$)
at the waveguide wall and using the wave equation Eq.~(\ref{variational:eq2}), we are led to the following eigenvalue problem:

\medskip
\emph{For a given value of $q$, find $\Omega \in \mathbb{C}$ and $\bu \in \mathcal{U}$ such that $\bu \neq \bm{0}$ and
$\forall \bm{v} \in \mathcal{U}$}
\begin{eqnarray}
\label{variational:eq3}
\int_{A}
(\nabla_S \bm{v} )^* : \left( \bm{c} : \nabla_S \, \bu \right) \, \dA 
=
\Omega^2 \, \int_{A} \rho \, \bm{v}^* \cdot \bu \, \dA  ,
\end{eqnarray}
where $A \subset \mathbb{R}^2$ is the cross-section of the waveguide
and $\mathcal{U}$ is the set of admissible solutions.

\medskip
For $q \neq 0$, we adopt the following change of variable
\begin{eqnarray}
\label{hat:u:z}
u_z & = & i \, q \, \hat{u}_z,
\end{eqnarray}
which simplifies the algebraic manipulation.
With the new variable $\hat{u}_z$, the symmetric gradient operator Eq.~(\ref{nabla:S:mode:eq1}) can be expressed as
\begin{eqnarray}
\label{nabla:S:mode:eq2}
\nabla_S \, \bu
=
\left[ \begin{array}{ccc}
\displaystyle
\frac{\partial }{\partial x} & 0 & 0 \\[12pt]
\displaystyle
0 & \displaystyle \frac{\partial }{\partial y} & 0\\[12pt]
\displaystyle
0 & 0 & \displaystyle - q^2 \\[12pt]
\displaystyle
0 & \displaystyle i \, q & \displaystyle i \, q \frac{\partial }{\partial y} \\[12pt]
\displaystyle
i \, q & 0 & i \, q \displaystyle \frac{\partial }{\partial x} \\[12pt]
\displaystyle
\frac{\partial }{\partial y} & \displaystyle \frac{\partial }{\partial x} & 0
\end{array}\right]
\,
\left[ \begin{array}{c}
u_x \\[12pt]  u_y \\[12pt]  \hat{u}_z
\end{array}\right] .
\end{eqnarray}
The variable $\hat{u}_z$ is particularly  useful for eigenproblems where the frequency $\Omega$
in Eq.~(\ref{variational:eq3}) is fixed to a given value while the propagation constant $q$ is the unknown eigenvalue.
Indeed, with the original displacement component $u_z$, this eigenproblem is nonlinear (as it involves both $q$ and $q^2$).
But, with the variable $\hat{u}_z$, the eigenproblem can be transformed into a linear eigenproblem where $q^2$ is the unknown eigenvalue,
if the propagation medium has a monoclinic symmetry, i.e., 
when the matrix $\bc$ has the profile shown in Eq.~(\ref{C:matrix:monoclinic}).
This result can be shown in a similar way to the derivation in~\cite{Dossou:CMAME:2005} for electromagnetic waveguides.

%
%

\subsection{Finite element calculation}

The numerical implementation of the finite element problem is similar to the one described in~\cite{Hladky1996}.
The finite element method is based on the approximation of the functional space $\mathcal{U}$
by a finite-dimensional space $\mathcal{U}_h \subset \mathcal{U}$.
When a set of basis functions $(\bm{g}_m)_{m=1,\dots, \text{ dim} \, (\mathcal{U}_h)}$ is chosen,
a field $\bm{g}_h$ inside the cross-section $A$ can be represented as
\begin{eqnarray}
\bm{g}_h = \sum_{m=1}^{{\rm dim \, (\mathcal{U}_h)}} u_m \, \bm{g}_m .
\end{eqnarray}
If we denote by $\bu_h$ the vector of expansion components $(u_m)_{m=1,\dots, {\rm dim \,} (\mathcal{U}_h)}$,
the application of classical finite element procedures to the continuous problem Eq.~(\ref{variational:eq3})
leads to the following finite dimensional eigenproblem:
\begin{eqnarray}
\label{variational:FEM}
\bm{K} \, \bu_h
=
\Omega^2 \, \bm{M} \, \bu_h ,
\end{eqnarray}
where, for $l,m \in \{ 1,\dots, {\rm dim \,} (\mathcal{U}_h) \}$, the elements of the matrices $\bm{K}$ and $\bm{M}$ are defined as
\begin{eqnarray}
K_{lm}
& = &
\int_{A}
(\nabla_S \bm{g}_l )^* : \left( \bc : \nabla_S \, \bm{g}_m \right) \, \dA , \\
M_{lm}
& = &
\int_{A} \rho \, \bm{g}_l^* \cdot \bm{g}_m \, \dA  .
\end{eqnarray}
For the computer implementation, we have used basis functions  $(\bm{g}_m)_{m=1,\dots, {\rm dim \,} (\mathcal{U}_h)}$ 
which are piecewise quadratic polynomials.
The generalised eigenvalue problem Eq.~(\ref{variational:FEM}) can be solved
using the eigensolver for sparse matrices ARPACK~\cite{ARPACK:SIAM:1998}.
ARPACK requires a linear system solver and we have relied on the
direct solver for sparse matrices UMFPACK~\cite{UMFPACK:ACM:2004}.
The finite element meshes were generated with the software Gmsh~\cite{Gmsh:IJNME:2009}.


\begin{thebibliography}{10}
\expandafter\ifx\csname url\endcsname\relax
  \def\url#1{\texttt{#1}}\fi
\expandafter\ifx\csname urlprefix\endcsname\relax\def\urlprefix{URL }\fi
\expandafter\ifx\csname href\endcsname\relax
  \def\href#1#2{#2} \def\path#1{#1}\fi

\bibitem{Eggleton2013}
B.~Eggleton, C.~Poulton, R.~Pant, {Inducing and harnessing stimulated
  {B}rillouin scattering in photonic integrated circuits}, Adv. Opt. Photonics
  5 (2013) 536--587.
\newblock \href {http://dx.doi.org/10.1364/AOP} {\path{doi:10.1364/AOP}}.

\bibitem{Rakich:PRX:2012}
P.~T. Rakich, C.~Reinke, R.~Camacho, P.~Davids, Z.~Wang, Giant enhancement of
  stimulated {B}rillouin scattering in the subwavelength limit, Phys. Rev. X 2
  (2012) 011008.
\newblock \href {http://dx.doi.org/10.1103/PhysRevX.2.011008}
  {\path{doi:10.1103/PhysRevX.2.011008}}.

\bibitem{VanLaer2015}
R.~V. Laer, A.~Bazin, B.~Kuyken, R.~Baets, D.~V. Thourhout, {Net on-chip
  {B}rillouin gain based on suspended silicon nanowires}, New J. Phys. 17~(11)
  (2015) 115005.
\newblock \href {http://dx.doi.org/10.1088/1367-2630/17/11/115005}
  {\path{doi:10.1088/1367-2630/17/11/115005}}.

\bibitem{Smith2017}
M.~J.~A. Smith, C.~M. {De Sterke}, C.~Wolff, M.~Lapine, C.~G. Poulton,
  {Enhanced acousto-optic properties in layered media}, Phys. Rev. B 96 (2017)
  064114:1--10.

\bibitem{Wolff2014}
C.~Wolff, M.~J. Steel, C.~G. Poulton, {Formal selection rules for {B}rillouin
  scattering in integrated waveguides and structured fibers}, Opt. Expr.
  22~(26) (2014) 32489.
\newblock \href {http://dx.doi.org/10.1364/OE.22.032489}
  {\path{doi:10.1364/OE.22.032489}}.

\bibitem{Wolff2015}
C.~Wolff, M.~J. Steel, B.~J. Eggleton, C.~G. Poulton, {Stimulated {B}rillouin
  scattering in integrated photonic waveguides: forces, scattering mechanisms
  and coupled mode analysis}, Phys. Rev. A 92 (2015) 013836:1--13.
\newblock \href {http://dx.doi.org/10.1364/OE.22.029270}
  {\path{doi:10.1364/OE.22.029270}}.

\bibitem{Florez:NC:2016}
O.~Florez, P.~F. Jarschel, Y.~A.~V. Espinel, C.~M.~B. Cordeiro, T.~P.~M.
  Alegre, G.~S. Wiederhecker, P.~Dainese, Brillouin scattering
  self-cancellation, Nat. Comm. 7 (2016) 11759.
\newblock \href {http://dx.doi.org/10.1038/ncomms11759}
  {\path{doi:10.1038/ncomms11759}}.

\bibitem{Sipe2016}
J.~E. Sipe, M.~J. Steel, {A {H}amiltonian treatment of stimulated {B}rillouin
  scattering in nanoscale integrated waveguides}, New J. Phys. 18 (2016)
  045004:1--21.
\newblock \href {http://dx.doi.org/10.1088/1367-2630/18/4/045004}
  {\path{doi:10.1088/1367-2630/18/4/045004}}.

\bibitem{numbatcode}
B.~C.~P. Sturmberg, K.~B. Dossou, M.~Smith, B.~Morrison, C.~Wolff, C.~Poulton,
  M.~J. Steel, Numbat---{T}he {N}umerical {B}rillouin {A}nalysis {T}ool,
  \url{https://github.com/bjornsturmberg/Numbat} (2018).

\bibitem{numbatdocs}
B.~C.~P. Sturmberg, K.~B. Dossou, M.~Smith, B.~Morrison, C.~Wolff, C.~Poulton,
  M.~J. Steel, Numbat---{T}he {N}umerical {B}rillouin {A}nalysis {T}ool,
  {D}ocumentation, \url{https://numbat-au.readthedocs.io/en/latest} (2018).

\bibitem{Dossou:CMAME:2005}
K.~Dossou, M.~Fontaine, A high order isoparametric finite element method for
  the computation of waveguide modes, Comput. Method. Appl. Mech. Eng.
  194~(6-8) (2005) 837--858.
\newblock \href {http://dx.doi.org/10.1016/j.cma.2004.06.011}
  {\path{doi:10.1016/j.cma.2004.06.011}}.

\bibitem{Sturmberg:CPC:2016}
B.~C. Sturmberg, K.~B. Dossou, F.~J. Lawrence, C.~G. Poulton, R.~C. McPhedran,
  C.~M. de~Sterke, L.~C. Botten, {EMUstack}: An open source route to insightful
  electromagnetic computation via the {B}loch mode scattering matrix method,
  Comput. Phys. Commun. 202 (2016) 276--286.
\newblock \href {http://dx.doi.org/10.1016/j.cpc.2015.12.022}
  {\path{doi:10.1016/j.cpc.2015.12.022}}.

\bibitem{Hladky:Hennion:JSV:1996}
A.-C. Hladky-Hennion, Finite element analysis of the propagation of acoustic
  waves in waveguides, J. Sound Vib. 194~(2) (1996) 119--136.
\newblock \href {http://dx.doi.org/10.1006/jsvi.1996.0349}
  {\path{doi:10.1006/jsvi.1996.0349}}.

\bibitem{Auld:book:1973}
B.~A. Auld, Acoustic Fields and Waves in Solids, Vol.~1, Wiley, 1973.

\bibitem{UMFPACK:ACM:2004}
T.~A. Davis, A column pre-ordering strategy for the unsymmetric-pattern
  multifrontal method, ACM Trans. Math. Softw. 30~(2) (2004) 165--195.
\newblock \href {http://dx.doi.org/10.1145/992200.992205}
  {\path{doi:10.1145/992200.992205}}.

\bibitem{Peterson:IJCSE:2009}
P.~Peterson, {F2PY}: a tool for connecting {F}ortran and {P}ython programs,
  Int. J. Comp. Sci. Eng. 4~(4) (2009) 296--305.
\newblock \href {http://dx.doi.org/10.1504/IJCSE.2009.029165}
  {\path{doi:10.1504/IJCSE.2009.029165}}.

\bibitem{Gmsh:IJNME:2009}
C.~Geuzaine, J.~F. Remacle, {Gmsh: a three-dimensional finite element mesh
  generator with built-in pre- and post-processing facilities}, Int. J. Numer.
  Meth. Eng. 71 (2009) 1309--1331.

\bibitem{VanLaer:NP:2015}
R.~V. Laer, B.~Kuyken, D.~V. Thourhout, R.~Baets, {Interaction between light
  and highly confined hypersound in a silicon photonic nanowire}, Nat.
  Photonics 9 (2015) 199--203.
\newblock \href {http://dx.doi.org/10.1038/nphoton.2015.11}
  {\path{doi:10.1038/nphoton.2015.11}}.

\bibitem{Kittlaus:2017}
E.~A. Kittlaus, N.~T. Otterstrom, P.~T. Rakich, On-chip inter-modal {B}rillouin
  scattering, Nat. Commun. 8 (2017) 15819:1--9.
\newblock \href {http://dx.doi.org/10.1038/ncomms15819}
  {\path{doi:10.1038/ncomms15819}}.

\bibitem{Morrison:Optica:2017}
B.~Morrison, A.~Casas-Bedoya, G.~Ren, K.~Vu, Y.~Liu, A.~Zarifi, T.~G. Nguyen,
  D.-Y. Choi, D.~Marpaung, S.~J. Madden, A.~Mitchell, B.~J. Eggleton, Compact
  {B}rillouin devices through hybrid integration on silicon, Optica 4~(8)
  (2017) 847--854.
\newblock \href {http://dx.doi.org/10.1364/OPTICA.4.000847}
  {\path{doi:10.1364/OPTICA.4.000847}}.

\bibitem{Smith:Wolff:OL:2016}
M.~J.~A. Smith, B.~T. Kuhlmey, C.~M. de~Sterke, C.~Wolff, M.~Lapine, C.~G.
  Poulton, Metamaterial control of stimulated {B}rillouin scattering, Opt.
  Lett. 41~(10) (2016) 2338--2341.
\newblock \href {http://dx.doi.org/10.1364/OL.41.002338}
  {\path{doi:10.1364/OL.41.002338}}.

\bibitem{Laude:AIP:2013}
V.~Laude, J.-C. Beugnot, Generation of phonons from electrostriction in
  small-core optical waveguides, AIP Adv. 3~(4) (2013) 042109.
\newblock \href {http://dx.doi.org/10.1063/1.4801936}
  {\path{doi:10.1063/1.4801936}}.

\bibitem{Choudhary:2017}
A.~Choudhary, B.~Morrison, I.~Aryanfar, S.~Shahnia, M.~Pagani, Y.~Liu, K.~Vu,
  S.~Madden, D.~Marpaung, B.~J. Eggleton, Advanced integrated microwave signal
  processing with giant on-chip {B}rillouin gain, J. Lightwave Technol. 35~(4)
  (2017) 846--854.
\newblock \href {http://dx.doi.org/10.1109/JLT.2016.2613558}
  {\path{doi:10.1109/JLT.2016.2613558}}.

\bibitem{Auld:book:1990}
B.~A. Auld, Acoustic Fields and Waves in Solids, 2nd Edition, Vol.~1, Krieger
  Publishing Co., 1990.

\bibitem{Hladky1996}
A.-C. Hladky-Hennion, Finite element analysis of the propagation of acoustic
  waves in waveguides, J. Sound Vib. 194 (1996) 119136.

\bibitem{ARPACK:SIAM:1998}
R.~B. Lehoucq, D.~C. Sorensen, C.~Yang, ARPACK users' guide - solution of
  large-scale eigenvalue problems with implicitly restarted Arnoldi methods,
  Software, environments, tools, SIAM, 1998.

\end{thebibliography}

\end{document}